\pdfoutput=1
\documentclass[prl,twocolumn,amsmath,amssymb,aps,superscriptaddress]{revtex4-2}

\usepackage[separate-uncertainty=true, multi-part-units=single]{siunitx}
\usepackage{xspace}
\usepackage{newtxtext, newtxmath}

\usepackage{graphicx}

\usepackage{pdfpages}

\makeatletter
\AtBeginDocument{\let\LS@rot\@undefined}
\makeatother

\usepackage{lineno}
\newcommand*\patchAmsMathEnvironmentForLineno[1]{
  \expandafter\let\csname old#1\expandafter\endcsname\csname #1\endcsname
  \expandafter\let\csname oldend#1\expandafter\endcsname\csname end#1\endcsname
  \renewenvironment{#1}
     {\linenomath\csname old#1\endcsname}
     {\csname oldend#1\endcsname\endlinenomath}}
\newcommand*\patchBothAmsMathEnvironmentsForLineno[1]{
  \patchAmsMathEnvironmentForLineno{#1}
  \patchAmsMathEnvironmentForLineno{#1*}}
\AtBeginDocument{
\patchBothAmsMathEnvironmentsForLineno{equation}
\patchBothAmsMathEnvironmentsForLineno{align}
\patchBothAmsMathEnvironmentsForLineno{flalign}
\patchBothAmsMathEnvironmentsForLineno{alignat}
\patchBothAmsMathEnvironmentsForLineno{gather}
\patchBothAmsMathEnvironmentsForLineno{multline}
}

\begin{document}

\title{Optimal rectification without forward-current suppression by biological molecular motor}

\author{Yohei Nakayama}
\affiliation{Department of Applied Physics, Graduate School of Engineering, Tohoku University, Aoba 6-6-05, Sendai 980-8579, Japan}
\author{Shoichi Toyabe}
\email{toyabe@tohoku.ac.jp}
\affiliation{Department of Applied Physics, Graduate School of Engineering, Tohoku University, Aoba 6-6-05, Sendai 980-8579, Japan}

\date{\today}
 
\newcommand{\Fone}{F\textsubscript{1}\xspace}
\newcommand{\Fo}{F\textsubscript{o}\xspace}
\newcommand{\FoFone}{F\textsubscript{o}F\textsubscript{1}\xspace}
\newcommand{\FF}{\FoFone}
\newcommand{\PP}{P\textsubscript{i}\xspace}

\newcommand{\Nex}{N_\mathrm{ex}}
\newcommand{\tauR}{\tau_\mathrm{rot}}
\newcommand{\tauI}{\tau_\mathrm{inh}}

\newcommand{\vbare}{v_\mathrm{bare}}
\newcommand{\vnet}{v_\mathrm{net}}

\newcommand{\Stradeoff}{S1\xspace}
\newcommand{\Srotationrate}{S2\xspace}
\newcommand{\Srotationrateparameter}{S3\xspace}
\newcommand{\Sduration}{S4\xspace}
\newcommand{\Sdurationdistribution}{S5\xspace}
\newcommand{\Spotential}{S6\xspace}
\newcommand{\Sdurationfitting}{S7\xspace}

\newcommand{\para}[1]{\medskip\par{\em #1}\/.---}

\begin{abstract}
We experimentally showed that biological molecular motor \Fone-ATPase (\Fone) implements an optimal rectification mechanism.
\Fone hardly suppresses adenosine triphosphate (ATP) synthesis, which is the \Fone's physiological role while inhibiting unfavorable hydrolysis of ATP.
This optimal rectification is a high contrast to a simple ratchet model, where
the inhibition of the backward current is inevitably accompanied by the suppression of the forward current.
The detailed analysis of single-molecule trajectories demonstrated a novel but simple rectification mechanism of \Fone with parallel landscapes and asymmetric transition rates.
\end{abstract}

\maketitle

\begin{figure}
 \centering
 \includegraphics{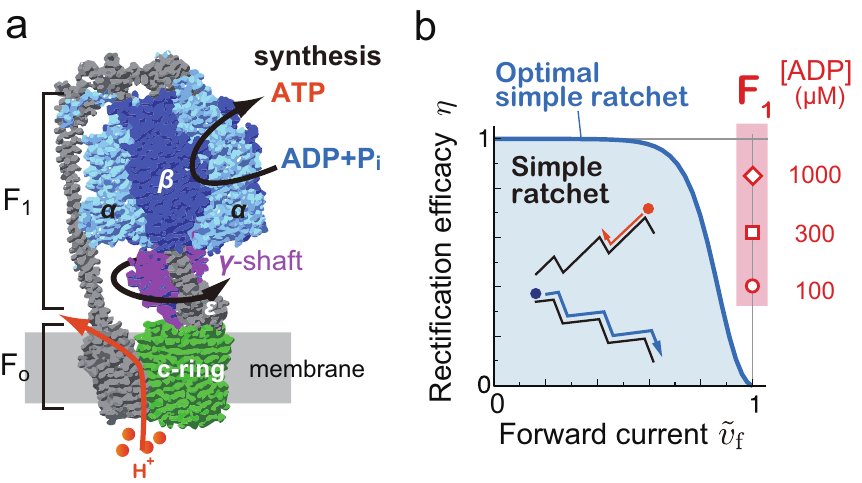}
 \caption{\FoFone-ATP synthase and current rectification.
 {\bf a}, \FoFone-ATP synthase converts the electrochemical potential of H\textsuperscript{+} to the chemical free energy of ATP via rotary machinery \cite{Guo2019}.
{\bf b}, The forward current vs. the rectification efficacy of \Fone and simple ratchet model.
The symbols correspond to the experimental results of \Fone under different concentrations of ADP. \([\mathrm{ATP}] = \SI{100}{\micro M}\) and \([\mathrm{\PP}] = \SI{1}{mM}\).
For \([\mathrm{ADP}] = \SI{1}{mM}\), we measured \(\tilde v\) only in \(\Nex = 0\) and assumed \(\tilde v_\mathrm{f} = 1\) as in the other cases.
The range of the current and rectification efficacy numerically obtained for the simple ratchet model with a saw-tooth potential is shown by the shaded area under the curve.
See Materials and Methods, and Fig.~\Stradeoff \cite{Note1} for details.
}
\label{fig:concept}
\end{figure}

\para{Introduction}
Energy transduction is essential for living systems.
In particular, adenosine triphosphate (ATP) plays a central role in biological energy transduction as a source of free energy for diverse processes.
The \FoFone is a cellular ATP factory \cite{Boyer1997} and composed of two coupled motors, \Fo and \Fone \cite{Abrahams1994,boyer1993,Guo2019} (Fig.~\ref{fig:concept}a).
The \Fo, driven by the flow of hydrogen ion (H$^+$) through it, rotates the $\gamma$-shaft of \Fone forcedly.
Then, \Fone synthesizes ATP from adenosine diphosphate (ADP) and inorganic phosphate (\PP) by converting the mechanical work to chemical free energy \cite{Itoh2004,Rondelez2005} with high efficiency \cite{Toyabe2011,Saita2015,Soga2017}.
However, there might be certain freely diffusing \Fone molecules that are not involved in the \FoFone complex.
Since \Fone is a reversible motor, these isolated \Fone rotate in the opposite direction and hydrolyze ATP to ADP and \PP.
Such futile ATP hydrolysis may also take place in the \FoFone complex when the driving force of \Fo becomes insufficient for synthesizing ATP.
 
The suppression of futile ATP consumption may be regarded as a rectification that blocks the unfavorable current (ATP hydrolysis) with sustaining the favorable current (ATP synthesis).
The rectification of current in small fluctuating systems such as biological molecular motors is not straightforward, since they work at an energy scale comparable to that of thermal energy \cite{Bustamante2005}.
For example, let us consider a Brownian particle in an asymmetric ratchet potential (Fig.~\ref{fig:concept}b), where the current responds asymmetrically to the reversal of the external field.
However, this simple ratchet model inherits a trade-off; the barrier needs to be high to block the backward current activated by thermal fluctuation, while the high barrier reduces the forward current as well.
We emphasize that the rectification discussed here is the asymmetry of the response to the reversal of the stationary driving force and is not that in the Brownian ratchet, which induces unidirectional current  by rectifying periodic or stochastic variation of the mechanical potential, driving force, or temperature \cite{astumian1997,Reimann2002}.

According to the literature, previous biochemical experiments imply that the phenomenon called inhibition serves as the rectification mechanism of \Fone for suppressing the futile ATP consumption  \cite{syroeshkin-atp-1995,bald-atp-1998,galkin-energy-dependent-1999}.
Besides, single-molecule experiments have shown that the inhibition causes long pauses of the rotation and can be activated mechanically \cite{Noji1997,HironoHara2001, Hirono-Hara2005}.
However, it remains elusive how the inhibition causes such a rectification, which has not been observed at the single-molecule level.
In this Letter, we aim to demonstrate the rectification by inhibition and reveal its mechanism. 
We have used single-molecule experiments of the isolated \Fone under external torque as a model system of \Fone working in the \FoFone complex.

\begin{figure}
 \centering
 \includegraphics{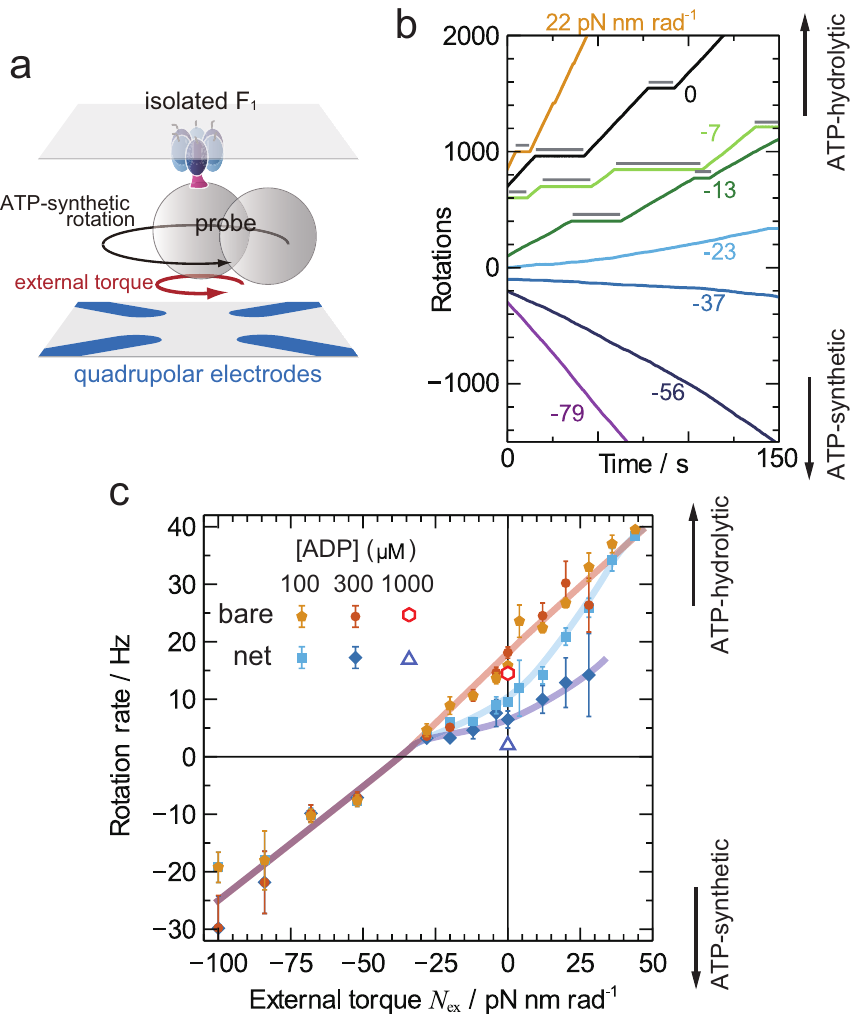}
\caption{Rotation of a single molecule of \Fone ($\alpha_3\beta_3\gamma$ subcomplex) under an external torque as a model system of \FoFone.
\textbf{a}, The schematics of the experimental setup. The rotation is probed by a \SI{300}{nm} dimeric probe particle attached to the \(\gamma\) shaft.
The electrorotation method using a 10-MHz rotating electric field induces external torque on the probe.
\textbf{b}, Single-molecule trajectories in the absence or presence of external torque with indicated values.
The gray lines indicate the pauses corresponding to the inhibited state.
The direction of the ATP-hydrolytic rotation is set as positive.
\textbf{c}, Torque dependence of \(\vbare\) and \(\vnet\) for different concentrations of ADP.
\([\mathrm{ATP}] = \SI{100}{\micro M}\) and \([\mathrm{\PP}] = \SI{1}{mM}\).
The rotation rates were averaged within bins with widths of \(8\) and \(16\) \si{pN nm / rad} for the ATP-hydrolytic and synthetic rotations, respectively.
The data at \(\Nex=0\) were averaged separately.
Data points are lacking near the stalled state because it was difficult to identify the inhibition pauses in this region where the rotation rate almost vanishes.
The fitting of the solid curves is a guide for the eye.
Error bars denote standard errors of the mean.
See Figs.~\Srotationrate and \Srotationrateparameter \cite{Note1} for the other condition, \([\mathrm{ATP}] = \SI{100}{\micro M}\), \([\mathrm{ADP}] = \SI{100}{\micro M}\), and \([\mathrm{\PP}] = \SI{100}{\micro M}\).
}
\label{fig:rotation}
\end{figure}

\para{Results}
We observed the rotation of a single \Fone molecule under an external torque, which is induced by the electrorotation method
to the probe fixed on the \(\gamma\) shaft \cite{Watanabe-Nakayama2008,Toyabe2010PRL,Toyabe2011} (Fig.~\ref{fig:rotation}a).
See Materials and Methods \footnote{See the Supplemental Material at [URL] for additional text and Figs~S1--S9.} for details of the experimental setup.
The rotation rate changes depending on the value of the external torque \(\Nex\).
When we applied a strong torque, the rotational direction was inverted to the ATP synthetic direction (Fig.~\ref{fig:rotation}b).

We observed intermittent long pauses of \Fone's rotation, as indicated by the gray lines in Fig.~\ref{fig:rotation}b.
The pauses occurred stochastically at three angular positions separated by 120$^\circ$, which is a characteristic of the inhibition \cite{HironoHara2001}.
We found that the inhibitory pauses took place only in the ATP-hydrolytic rotations and not in the ATP-synthetic rotations.
This selective inhibition of the ATP hydrolysis is consistent with the previous bulk experiments \cite{syroeshkin-atp-1995,bald-atp-1998,galkin-energy-dependent-1999}, supporting the idea that the inhibition serves as the rectification mechanism.

To characterize the inhibition, the rotational trajectories were divided into the rotating and inhibited states based on the instantaneous rotation rates (see Materials and Methods \cite{Note1} for details).
Figure~\ref{fig:rotation}c and Fig.~\Srotationrate in \cite{Note1} show
the bare and the net rotation rates, \(\vbare\) and \(\vnet\), corresponding to the rotation excluding and including the inhibition, respectively.
As already implied by the trajectories in Fig.~\ref{fig:rotation}b, the inhibition-mediated suppression of the net rotation rate was only seen in the ATP-hydrolytic rotation.
The increase in the ADP concentration further suppressed \(\vnet\).
On the other hand, \(\vbare\) was not significantly affected.

We here introduce the mean normalized current \(\tilde v\) and the rectification efficacy \(\eta\) for evaluating the performance of the rectification. 
\(\tilde v\) is the ratio between the mean currents with and without the rectification mechanism and is given as \(\tilde v = v_\mathrm{net} / v_\mathrm{bare}\) for \Fone.
\(\eta\) is defined based on the response asymmetry as \(\eta = 1 - |\tilde v(N)| / |\tilde v(-N)|\), where \(N\) is the driving force.
Since \Fone hydrolyzes three ATP molecules per rotation, the driving force by \Fone is given as the free energy change associated with an ATP hydrolysis \(\Delta\mu\) divided by \SI{120}{\degree}, that is \(N_\mathrm{motor} = \Delta\mu / \SI{120}{\degree}\).
Then, \(N = \Nex + N_\mathrm{motor}\).
The values of \(\Delta\mu\) are calculated by a method developed in \cite{Krab1992} and are in the range of \SI{63}{pN nm} to \SI{82}{pN nm} in our experimental conditions.
The relation between the forward normalized current $\tilde v_\mathrm{f}=|\tilde v(-N)|$ and $\eta$ at $N=N_\mathrm{motor}$ of \Fone is compared with that of a simple ratchet model in Fig.~\ref{fig:concept}b (see Materials and Methods \cite{Note1} for details and Fig.~\Stradeoff for the other values of \(N\)).
We observed that the \Fone achieved large $\eta$ values without suppressing the ATP-synthetic current (\(\tilde v_\mathrm{f} \simeq 1\)), whereas the simple ratchet model could not reach \(\eta > 0\) and \(\tilde v_\mathrm{f} = 1\) simultaneously.
Thus, \Fone implements a rectification mechanism that circumvents the rectification trade-off  between \(\tilde v_\mathrm{f}\) and \(\eta\) inherited by the simple ratchet.

\begin{figure}[b]
 \centering
 \includegraphics{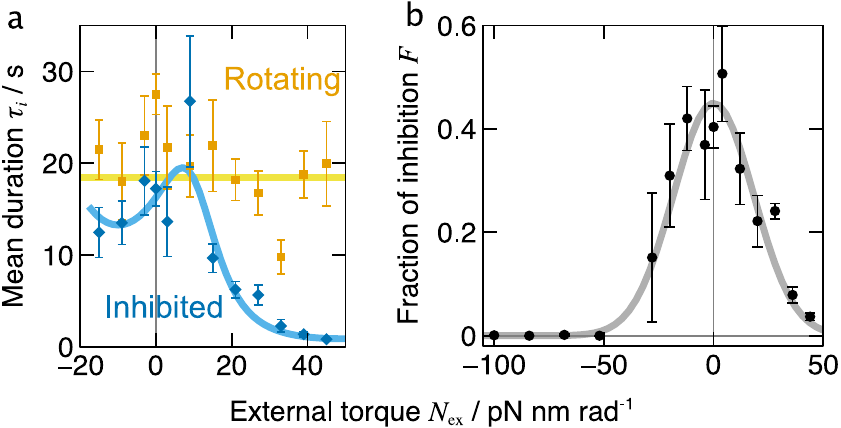}
\caption{Torque dependence of \(\tauR\) (yellow square) and \(\tauI\) (blue diamond) ({\bf a}), and \(F\) ({\bf b}).
\([\mathrm{ATP}] = \SI{100}{\micro M}\), \([\mathrm{ADP}] = \SI{100}{\micro M}\), and \([\mathrm{\PP}] = \SI{1}{mM}\).
The blue curve in {\bf a} is the fitting curve to \(\tauI\) by Eq.~(\ref{e:tauI_}) and (\ref{e:rho_}). The obtained values of the fitting parameters are \(\theta^*-L = \SI{12}{\degree}\) and \(S - \theta^* = \SI{59}{\degree}\).
The yellow line is the mean of \(\tauR\).
The gray curve in {\bf b} shows the fitting by a Gaussian function.
The widths of the bins for averaging are
\(\SI{6}{pN nm / rad}\) for \(\tauR\) and \(\tauI\),
and
the same as Fig.~\ref{fig:rotation}c for \(F\).
The plots of \(\tauR\) and \(\tauI\) are limited for \(\Nex > \SI{-20}{pN nm / rad}\) because no pause is observed otherwise.
Error bars are standard errors of the mean.
See Fig.~\Sduration for the other conditions and Fig.~\Sdurationdistribution for the distributions of the duration of each state \cite{Note1}.
}
\label{fig:duration}
\end{figure}

Next, we examined the mean duration of the rotating and inhibited states to evaluate the inhibition dynamics.
The mean duration of the inhibited state,  \(\tauI\), exhibited a peak at $\Nex\simeq 0$ (Fig.~\ref{fig:duration}a).
On the other hand, the mean duration of the rotating state, \(\tauR\), had no significant dependence on \(\Nex\).
Accordingly, the time fraction of the inhibited state \(F=\tauI/(\tauI+\tauR)\) has a peak at \(\Nex\simeq 0\)  (Fig.~\ref{fig:duration}b). 
Hence we concluded that the suppression of the rotation is mainly regulated by \(\tauI\).
An important feature is that the external torque in both directions reduced the value of \(\tauI\).
This observation implied the existence of two activation paths from the inhibited state, given the previous observations that \Fone's elementary reactions typically depend on the angle monotonically \cite{watanabe-mechanical-2012}.

\begin{figure}
 \centering
 \includegraphics[bb = 0 0 209 233]{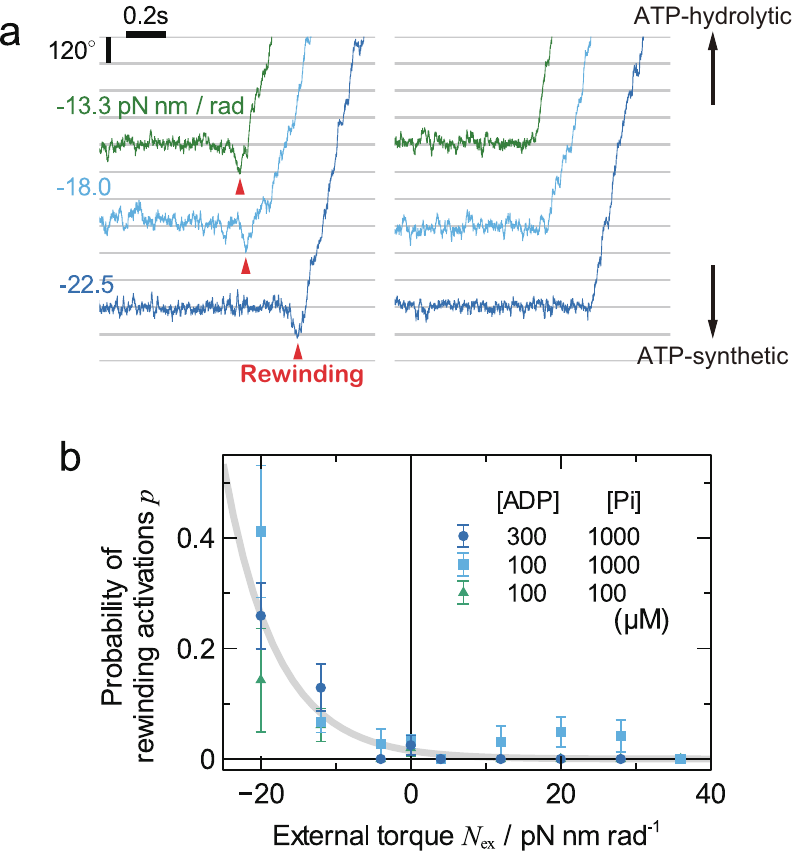}
\caption{Dynamics of activation from the inhibited state.
\textbf{a}, Trajectories of activation with (left) or without (right) the transient rewinding in the ATP-synthetic direction (red triangles) under indicated values of external torque.
\([\mathrm{ATP}] = \SI{100}{\micro M}\), \([\mathrm{ADP}] = \SI{100}{\micro M}\), and \([\mathrm{\PP}] = \SI{1}{mM}\) for all trajectories.
\textbf{b}, Probability of activation with rewinding \(p\) at different values of torque for various concentrations of ADP and \PP. 
\([\mathrm{ATP}] = \SI{100}{\micro M}\).
We did not estimate \(p\) for \(\Nex < \SI{-20}{pN nm / rad}\) as inhibition was rarely observed in that region.
The widths of bins are \SI{8}{pN nm / rad}.
The solid curve is an exponential function as a guide to the eye.
See \cite{Note1} for error bars.
}
\label{fig:activation}
\end{figure}

The activation dynamics observed in the rotational trajectories indicated the involvement of two such activation paths (Fig.~\ref{fig:activation}).
We found that \Fone is stochastically activated in two ways: activation with or without a transient rewinding in the ATP-synthetic direction (Fig.~\ref{fig:activation}a).
The magnitude of rewinding reached up to \(\sim \SI{120}{\degree}\).
The probability of the rewinding activation, \(p\), increased when an external torque was applied in the ATP-synthetic direction (Fig.~\ref{fig:activation}b).
This result is consistent with the picture that there are two elementary reactions whose rates monotonically depend on the angle.
The qualitative tendency was the same for different concentrations of ADP and \PP.
Note that \(p\) may be underestimated since it is not always possible to identify the short rewinding rotations by discriminating them from the thermal fluctuations.

\begin{figure}[t]
\begin{center}
\includegraphics[]{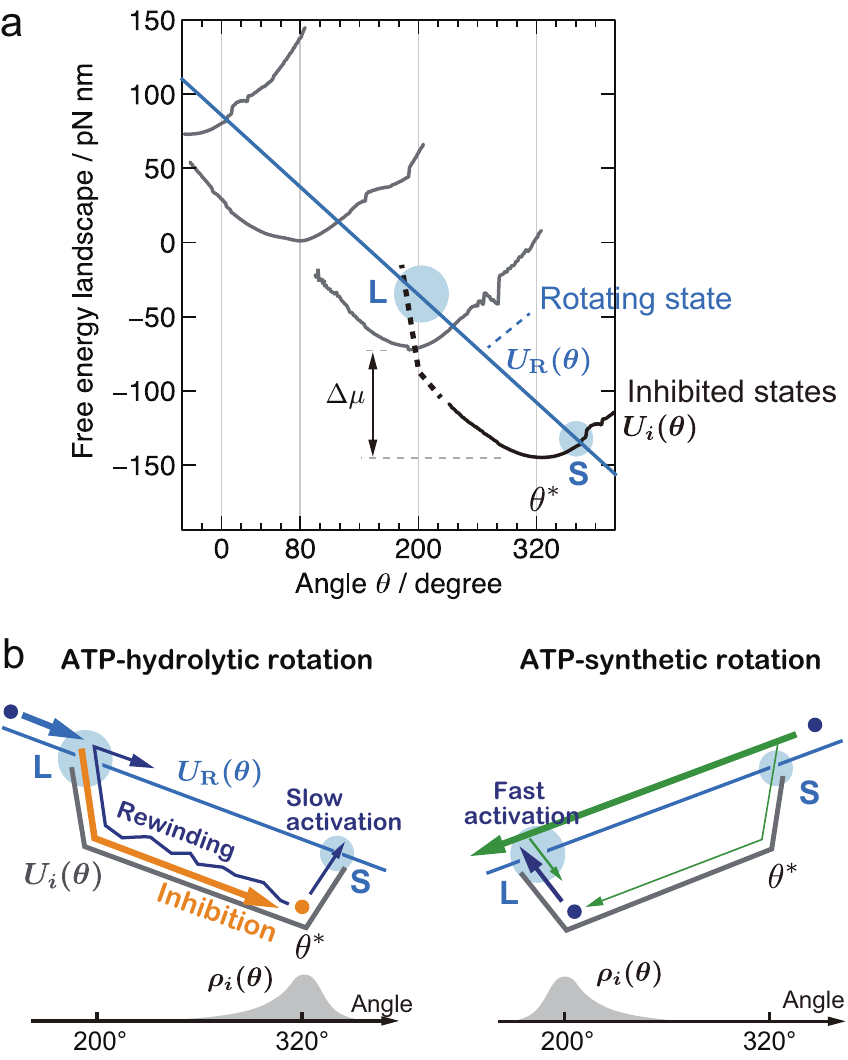}
\caption{Model depicting the rectification by \Fone.
{\bf a}, Free energy landscapes of the inhibited states \(U_i(\theta)\) recovered from the trajectories of a single \Fone molecule. 
The dashed part is a guide for the eye to indicate the correspondence with Fig.~\ref{fig:scheme}b.
Three free energy landscapes correspond to three inhibited states separated by \SI{120}{\degree}. 
The landscape minima are vertically shifted by \(\Delta\mu = \SI{73}{pN nm}\) from each other.
The roughly estimated free energy landscape of the rotating state \(U_\mathrm{R}(\theta)\) is shown by the blue line.
 \([\mathrm{ATP}] = \SI{100}{\micro M}\),
 \([\mathrm{ADP}] = \SI{100}{\micro M}\),
 and \([\mathrm{\PP}] = \SI{1}{mM}\).
See \cite{Note1} for the definition of the angle origin and the relative height between \(U_R(\theta)\) and \(U_i(\theta)\).
\textbf{b}, Our proposed schematic model of rectification by \Fone.
In the ATP-hydrolytic rotation, the inhibited state serves as a trap (``Inhibition'').
The torque to the ATP-synthetic direction enhances the activation with the rewinding to \(L\) (``Rewinding'').
In contrast, the inhibited state is quickly activated in the ATP-synthetic rotation (``Fast activation'').
}
\label{fig:scheme}
\end{center}
\end{figure}

Finally, we recovered the free energy landscapes corresponding to the three inhibited states separated by \SI{120}{\degree} (Figs.~\ref{fig:scheme}a and \Spotential in \cite{Note1}) to model the rectification mechanism.
The \(i\)-th landscape \(U_i(\theta)\) was recovered from the angular distributions in the corresponding inhibited state for multiple values of the external torques (see Materials and Methods \cite{Note1} for details).
In addition, we roughly estimated the free energy landscape of the rotating state \(U_\mathrm{R}(\theta)\) as a straight line with a slope of \(N_\mathrm{motor}\).
We found that the left side of \(U_i(\theta)\) has a slope close to $N_\mathrm{motor}$ and is nearly parallel to \(U_\mathrm{R}(\theta)\).
Since the reversal of the driving force changes these slopes from \(N_\mathrm{motor}\) to \(-N_\mathrm{motor}\),
the free energy landscape has a mirror symmetry between the ATP-hydrolytic and -synthetic rotations.

\para{Model}
\newcommand{\alphaL}{\alpha_L}
\newcommand{\alphaS}{\alpha_S}
Based on the experimental results, we propose a model for describing the rectification mechanism  (Fig.~\ref{fig:scheme}b).
The rectification is the asymmetric response to the reversal of the driving force.
The presence of the symmetry of the free energy landscape indicates that the rectification does not arise from the energetic asymmetry but from the kinetic asymmetry between the activation rates.
We here validate this kinetic asymmetry.

The experimental results suggested two paths for the activation from the inhibited state to the rotating state.
Let $L$ and $S$ be the locations of the transition from the inhibited state to the rotating state (Fig.~\ref{fig:scheme}b).
In this model, \(\tauI\) is given as
\begin{align}
 \tauI = [\alphaL \rho_i(L) + \alphaS \rho_i(S)]^{-1},
 \label{e:tauI_}
\end{align}
where \(\rho_i(\theta)\) is the probability density function of \(\theta\) in the \(i\)-th inhibited state, and \(\alphaL\) and \(\alphaS\) are the constant parameters.
The result that $\tauI$ has a peak at $\Nex\simeq 0$ (Fig.~\ref{fig:duration}a) implies that the activation rates at $L$ and $S$, that is, $\alpha_\mathrm{L}\rho_i(L)$ and $\alpha_\mathrm{S}\rho_i(S)$, have similar values at $\Nex\simeq 0$.
On the other hand,  it is natural to expect $\rho_i(L)\ll \rho_i(S)$ since \(U_i(L)>U_i(S)\) follows from the locations of \(L\) and \(S\), which are \(\SI{120}{\degree}\) backward \cite{Watanabe2010} and \(\sim\SI{40}{\degree}\) forward \cite{HironoHara2001} from \(\theta^* = \mathop\mathrm{argmin}_\theta U_i(\theta)\), respectively.
Thus, a large kinetic asymmetry, $\alpha_L/\alpha_S\gg 1$, is deduced.
See \cite{Note1} for the validation of the above argument based on the reaction scheme.

We next quantitatively evaluate the value $\alpha_L/\alpha_S$ from the dependence of \(\tauI\) on \(\Nex\) (Fig.~\ref{fig:duration}).
Assuming that the angular distribution is sufficiently equilibrated in the inhibited state, $\rho_i(\theta) = e^{ -\beta(U_i(\theta) - \Nex \theta)}/Z$ holds, where  \(\beta\) is the ambient inverse temperature, and \(Z = \int e^{ -\beta(U_i(\theta) - \Nex \theta)} \mathrm d\theta\) is the partition function.
By evaluating \(Z\) using the saddle point approximation, \(\rho_i(\theta)\) is rewritten as
\begin{align}
 \rho_i(\theta) \propto \exp{\left\{-\beta\left[U_i(\theta) - U_i(\theta^*) - \Nex(\theta - \theta^*) + \frac{\Nex^2}{2\kappa}\right]\right\}},
 \label{e:rho_}
\end{align}
where the curvature \(\kappa = U_i''(\theta)\) is regarded as a constant.
Equation~(\ref{e:tauI_}) with Eq.~(\ref{e:rho_}) fitted $\tauI$ well within the experimental errors (Fig.~\ref{fig:duration}a).
The fitting parameters gives a ratio \( \alphaL e^{-\beta U_i(L)} / \left(\alphaS e^{-\beta U_i(S)}\right) = 19\).
Given that \(U_i(L) - U_i(S) \sim \Delta\mu\), 
we obtain a rough estimation of the kinetic asymmetry as \(\alphaL / \alphaS \sim 10^8\).
The angular equilibration assumed here might not apply to \(L\) because of its fast kinetics.
However,  \(\alphaL / \alphaS\) becomes larger in this diffusion-limited situation.
Thus, the above evaluation provides the lower bound of the asymmetry.

The enormous kinetic asymmetry between \(L\) and \(S\) brings about the rectification in the following manner.
If the free energy landscape is completely symmetric as illustrated in Fig.~\ref{fig:scheme}b,
the local detailed balance condition \cite{Seifert2012} imposes that the ratio of the transition rates between the rotating and inhibited states at $L$ is the same as that at $S$. 
Thus, the result $\alphaL/\alphaS\gg 1$ indicates a significantly greater transition rate from the rotating to the inhibited state at \(L\) than that at \(S\).
Therefore, in the absence of external torque, the ATP-hydrolytic rotation is easily trapped by the inhibited state through \(L\) and is hardly activated since the activation rate at \(S\) is small and the thermal diffusion to \(L\) is rare.
On the other hand, in the ATP-synthetic rotation, a rapid activation takes place through \(L\), which is close to the free energy minimum of the inhibited state.
Hence, we conclude that the rectification arises from the kinetic asymmetry.
Note that this rectification mechanism itself does not consume free energy.

\para{Concluding remarks}
We showed that the core mechanism of rectification by \Fone is derived from the kinetic asymmetry between the two activation paths from the inhibited state (Fig.~\ref{fig:scheme}b).
This mechanism circumvents the trade-off inherited by the simple ratchet; a high barrier for the brake of the backward current inevitably suppresses the forward current as well.
A minimal modification to circumvent this trade-off is to add a landscape that serves as a brake.
The large efficacy without the suppression of the forward current is achieved by changing the probability of being in the additional landscape.
Our striking finding is that \Fone employs kinetic asymmetry instead of the asymmetry of the free energy landscape to change the probability.
Current control is ubiquitous across biological systems, including voltage-gated ion channels.
It is intriguing whether the above mechanism is common to biological systems or is a characteristic of the \Fone.

The development of a theoretical aspect may help elucidate the design principle of the rectification mechanism.
Current is a fundamental quantity in nonequilibrium physics.
General frameworks such as the fluctuation theorem \cite{Seifert2012} and thermodynamic uncertainty relation \cite{barato2015,dechant2020} have been developed to characterize current.
However, to the best of our knowledge, the general framework of current rectification has not yet been investigated.
The development of a theoretical framework would provide a unified viewpoint for these processes.

The refinement of the model is also an essential task for better quantitative evaluation.
For example, Eqs.~(\ref{e:tauI_}) and (\ref{e:rho_}) with fixing the angular distance \(\theta^* - L = \SI{120}{\degree}\) cannot fit \(\tauI\) in the range of \(\Nex < 0\) (Fig.~\Sdurationfitting in \cite{Note1}), whereas \(\theta^* - L\) is expected to be equal to the maximum magnitude of the transient rewinding, \SI{120}{\degree} (Fig.~\ref{fig:activation}a).
In addition, such a large transient rewinding may not be expected to be driven by rotational Brownian motion on the recovered free energy landscape (Fig.~\ref{fig:scheme}a).
These may imply that there are multiple chemical states in the inhibited state.
In such a case, it would be necessary to analyse the chemical state-specific landscapes as done in \cite{Toyabe2012} and reexamine the values of the angular distances, \(\theta^* - L\) and \(S - \theta^*\).
Therefore, we expect our results to be the first step towards a detailed analysis of the inhibited state.

We thank Eiro Muneyuki for technical assistance with the sample preparation.
We appreciate the helpful discussions with Yuki Izumida and Tomoaki Okaniwa.
This work was supported by JSPS KAKENHI (JP18H05427, JP19H01864).

\onecolumngrid

\newpage 

\section*{Supplementary material}

# Supplemental Material for "Optimal rectification without forward-current suppression by biological molecular motor"

Yohei Nakayama[1] and Shoichi Toyabe[1]

[1]*Department of Applied Physics, Graduate School of Engineering,*
*Tohoku University, Aoba 6-6-05, Sendai 980-8579, Japan*

## I. MATERIALS AND METHODS

### A. Preparation of $F_1$

We used the $\alpha_3\beta_3\gamma$ subcomplex of $F_1$ derived from a thermophilic *Bacillus* PS3 with mutations for the rotation assay (His$_6$-$\alpha$C193S/W463F, His$_{10}$-$\beta$, and $\gamma$S107C/I210C) [1]. *E. coli* strain JM103 $\Delta$unc carrying the expression plasmid of this $F_1$ was cultured in 300 ml of Terrific broth containing 50 μg/ml ampicillin at 37 °C overnight. The cells collected by centrifugation for 15 min at $9.3 \times 10^3 g$ were resuspended in 50 mM phosphate buffer at pH 8.0 containing 20 mM imidazole and 300 mM potassium chloride (buffer I). The suspended cells were sonicated for 20 min by ultrasonic disrupter (TOMY SEIKO, Japan) to be disrupted. The cell lysate was incubated at 60 °C for 20 min, and centrifuged for 25 min at $1.9\times10^4 g$. The supernatant was applied to a Ni-NTA Superflow column (QIAGEN, Germany) equilibrated with buffer I. After washing the column with buffer I, $F_1$ were eluted by 50 mM phosphate buffer at pH 8.0 containing 250 mM imidazole and 300 mM potassium chloride. A reducing agent, ($\pm$)-dithiothreitol, was added to the solution of the eluted $F_1$ at a final concentration of 1.5 mM, and reacted for 1 h at room temperature. The solution was concentrated by a centrifugal filter device (Merck, Germany), and applied to a Superdex™ 200 Increase column (GE Healthcare Life Sciences, IL) equilibrated with 100 mM potassium buffer at pH 7.0 containing 2 mM ethylenediaminetetraacetic acid. The concentration of $F_1$ was measured by spectrophotometer (JASCO, Japan), and the solution was reacted with biotin-PEAC$_5$-maleimide (purchased from Dojindo, Japan) at a molar ratio of 1:3 for 30 min. The solution was reacted with 0.5 μM ($\pm$)-dithiothreitol for 10 min at room temperature. The samples were flash-frozen in liquid nitrogen and stored at −80 °C.

### B. $Ni^{2+}$-NTA modification of cover glasses

In our single-molecule experiments, $F_1$ molecules were adhered to a cover glass modified by $Ni^{2+}$-NTA. Since the $Ni^{2+}$-NTA coating disturbed the electrorotation torque at high ADP concentration, we also used a KOH-treated cover glass in some experiments. We observed no significant difference in the data between these two cover glasses.

The $Ni^{2+}$-NTA glass was prepared as follows. Cover glasses ($24\times36\,mm^2$, thickness No. 1; Matsunami) were washed with ion-exchanged water three times and immersed in 10 M potassium hydroxide solution overnight. The glasses were washed with ion-exchanged water six times, and immersed in a 1:1 mixture of ethanol and ion-exchanged water containing 2 % v/v (3-Mercaptopropyl)trimethoxysilane (purchased from TCI, Japan) and 0.06 % v/v acetic acid for 120 min at 60 °C. They were washed with ion-exchanged water three times, and baked at 120 °C for 60 min. They were immersed in 50 mM phosphate buffer (pH 6.7) containing 1 mM ($\pm$)-dithiothreitol and 2 mM ethylenediaminetetraacetic acid for 60 min. They were washed with ion-exchanged water six times, and reacted with 1.8 mg/ml Maleimido-C$_3$-NTA (purchased from Dojindo, Japan) for 1 h in 100 mM phosphate buffer (pH 6.7). The glasses were immersed in 50 mM nickel sulfate solution for 30 min, and washed with ion-exchanged water, before using them for single-molecule experiments.

The KOH-treated cover glass was prepared by washing the glasses with ion-exchanged water six times after being immersed in 10 M potassium hydroxide solution overnight.

### C. Single-molecule experiments

The experimental setup is essentially the same as that in the previous studies [2]. An observation chamber consisted of a $Ni^{2+}$-NTA modified or KOH-treated cover glass and a slide with quadrupolar electrodes. The cover glass and the slide were separated by double-sided adhesive tape (10 μm thickness; Teraoka, 7070W) with silicone grease (Shin-Etsu Chemical, Japan) on a surface. The solution of $F_1$ was diluted with 50 mM MOPS buffer (pH 6.7) containing 5 mg/ml bovine serum albumin, 50 mM potassium chloride, and 1 mM magnesium chloride (buffer II) to a final concentration of 1 nM. The chamber was filled with the solution, and incubated for 10 min to immobilize $F_1$ molecules on the surfaces.

The bovine serum albumin was added as the blocking agent. The chamber was washed with buffer II, and streptavidin-coated polystyrene particles (diameter = 276 nm, Thermo Fisher Scientific) diluted with buffer II was infused into it. Azide contained in the solution of polystyrene particle was removed in advance by repeating centrifugation, exchange of supernatant, and re-dispersion six times. After a 30 min incubation, the solution in the chamber was exchanged with 5 mM MOPS buffer containing 1 mM magnesium chloride, 100 µM MgATP, and indicated amount of MgADP and $P_i$ (pH 7.0 at the room temperature).

Rotation of the $\gamma$-shaft was probed by dimeric polystyrene particles attached to the biotinylated $\gamma$-shaft. The observation was performed on a phase-contrast upright microscope (Olympus, Japan) with a 60× objective (NA1.42), high-intensity LED (623 nm, 4.8 W, Thorlabs, NJ) for illumination, a high-speed camera (Basler, Germany) at 4,000 Hz, and a laboratory-made capturing software developed on LabVIEW (National Instruments, TX). The angular position of the dimeric probe was analyzed by an algorithm based on a principal component analysis of the probe image. Because of the relatively high ATP concentration (100 µM) and the relatively large probe size (300 nm dimeric particle), we could not resolve the step-wise rotation corresponding to the elementary reactions such as the substrate binding, unbinding, and ATP hydrolysis/synthesis reactions.

The temperature in the chamber was kept at 37.9 ± 0.3 °C by heaters attached to the objective and embedded in a stage. The heating speeds up the reaction cycles and increases the data volume. The temperatures of the objective and the stage were monitored by thermocouples and controlled by PID controllers. The temperature in the chamber was measured in advance by using a thin thermocouple (Toa Electric, Japan) to calibrate the relation of it to the temperatures of the objective and the stage.

We used a laboratory-made autofocus system to keep the probe in focus since each observation lasts for five minutes. The contrast of the probe images was calculated in real-time, and a stepping motor (Oriental motor, Japan) kept rotating a focus knob to the direction that the contrast of the image increases.

The number of the trajectories we observed are 98 (9 molecules) for [ATP] = [ADP] = 100 µM and [$P_i$] = 1 mM, 69 (7 molecules) for [ATP] = 100 µM, [ADP] = 300 µM and [$P_i$] = 1 mM, 82 (8 molecules) for [ATP] = [ADP] = [$P_i$] = 100 µM, and 5 (5 molecules) for [ATP] = 100 µM and [ADP] = [$P_i$] = 1 mM.

### D. Electrorotation

We applied torque on the probe by using a rotating electric field at 10 MHz generated with the quadrupolar electrodes patterned on the glass surface of the chamber [3–6]. A 10-MHz sinusoidal voltage with a phase shift of $\pi/2$ was induced on the four electrodes. The distance between the electrodes is 47 µm, and the chamber height was about 20 µm. The signals generated by a function generator (nf, Japan) were divided by 180° phase divider (Thamway, Japan), amplified by four amplifiers (Analog Devices, MA), and loaded on the electrodes. This generates an electric field rotating at 10 MHz in the center of the electrodes and induces a dipole moment rotating at 10 MHz on the dimeric probe. Since there is a phase delay of the dipole moment with respect to the electric field, the dimeric probe is subjected to a constant torque. The torque magnitude was regulated by the multifunction board (National Instruments, TX) equipped on PC, which controls the voltage amplitude $V_0$ of the signals. The torque magnitude is proportional to the square of $V_0$ [5, 7]. The camera and amplitude signal were synchronized at a time difference of less than one microsecond. The method of the torque calibration is based on the fluctuation-dissipation theorem at the high-frequency region. We set 800 Hz as a frequency for the torque calibration. See [6] for the details of the torque calibration method.

### E. Identification of the rotating and inhibited states

We analyzed the rotational trajectories to identify the rotating and inhibited state based on the instantaneous rotation rates. Instead of analyzing the whole trajectories, we selected the frames every 200 frames to suppress the effect of the estimation error in the angular position. A hidden Markov model [8] was applied to the series of angular displacements between the successive selected frames for identifying the states. In our modeling, the latent variable $z$ represents the state (rotating or inhibited), the angular displacement is generated from Gaussian distributions whose mean and variance depend on $z$. The model parameters were optimized by the Baum-Welch algorithm, and the most probable sequence of $z$ was found by the Viterbi algorithm [8]. Then, we ignored inhibited states shorter than 300 ms, since a subtle fluctuation of the rotation rate, for example originated from the asymmetric adhesion of $F_1$ to a glass surface, may be detected as such short pauses. The change in this threshold or the interval between selected frames did not make a qualitative difference (Figs. S3 and S4). The trajectories which were not distinctly divided into the rotating and pausing states were manually excluded from the analysis.

### F. Current and rectification efficacy of the simple ratchet model

The motion of the Brownian particle in a static potential $V(x)$ is described by a Langevin equation

$$\Gamma \dot{x} = f - \frac{dV(x)}{dx} + \sqrt{2\Gamma\beta^{-1}}\xi, \tag{S1}$$

where $x$ is the position of the particle, $\Gamma$ is the frictional constant, $f$ is the driving force, $\beta$ is the ambient inverse temperature, and $\xi$ is a white Gaussian noise with zero mean and unit variance. The mean current $v(f)$ in the steady state under a driving force $f$ is expressed as [9]

$$v(f) = \frac{1 - \exp(-f)}{\int_0^1 dx \int_0^1 dy \exp\left(-V(x) + V(y+x) - fy\right)}. \tag{S2}$$

Here, we set $\beta$, $\Gamma$, and the period of $V(x)$ unity. We evaluated $v(f)$ and $v(-f)$ for

$$V(x) = \begin{cases} \Delta V \dfrac{x}{x_{\text{peak}}} & (0 \le x \le x_{\text{peak}}) \\ \Delta V \dfrac{(1-x)}{(1-x_{\text{peak}})} & (x_{\text{peak}} \le x \le 1) \end{cases} \tag{S3}$$

by the numerical integration of Eq. (S2). $\Delta V$ and $x_{\text{peak}}$ are the height and the position of the peak of the ratchet potential, respectively. The faster (slower) one of $|v(f)|$ and $|v(-f)|$ was defined as the mean forward (backward) currents. The currents were normalized by the current without the ratchet potential, which equals to $f$.

### G. Recovery of the free energy landscapes in the inhibited states

Each free energy landscape of the inhibited state $U_i(\theta)$ was recovered from the angular histograms $v_i(\theta; N_{\text{ex}}^{(k)})$ in the corresponding inhibited state for multiple values of the external torques $\{N_{\text{ex}}^{(k)}\}$. We assume that the angular degree of freedom is sufficiently equilibrated in each landscape, since $\tau_{\text{inh}}$ is much longer than the relaxation time in each landscape, which is of the order of $10\,\text{ms}$. Under this assumption, $v_i(\theta; N_{\text{ex}}^{(k)})$ should satisfy

$$v_i(\theta; N_{\text{ex}}^{(k)}) \propto \exp\left[-\beta\left(U_i(\theta) - N_{\text{ex}}^{(k)}\theta\right)\right]. \tag{S4}$$

Therefore, we sought $U_i(\theta)$ that minimizes a cost function

$$\sum_k v_i(\theta; N_{\text{ex}}^{(k)}) \cdot \left[U_i(\theta) - N_{\text{ex}}^{(k)}\theta + \beta^{-1}\ln v_i(\theta; N_{\text{ex}}^{(k)}) + u_i^{(k)}\right]^2, \tag{S5}$$

where $u_i^{(k)}$ are parameters optimized to adjust the origins of the free energy landscapes. The minimization of Eq. (S5) corresponds to the method of weighted least squares. Since the variance of $v_i(\theta; N_{\text{ex}}^{(k)})$ is expected to be proportional to itself, we chose $v_i(\theta; N_{\text{ex}}^{(k)})$ as the weight. See Fig. S6 for $v_i(\theta; N_{\text{ex}}^{(k)})$ and how free energy landscapes is recovered.

In Fig. 5a in the main text, we plotted $U_i(\theta)$ together with the roughly estimated free energy landscape of the rotating state $U_R(\theta)$. The relative height between $U_R(\theta)$ and $U_i(\theta)$ was determined based on the previous experimental result with magnetic tweezer [10]. Their experiment showed that the probability of activation at the fixed angle reaches $50\,\%$ at $40°$ forward from the bottom of $U_i(\theta)$. We consider that this result indicates $U_i(\theta) = U_R(\theta)$ at those angle, and the origin of $U_i(\theta)$ was set so as to reproduce this property. However, we should keep in mind that the saturation of the probability with the time was not verified in the presence of ADP in [10].

### H. Details of data analysis

The bare rotation rate, $v_{\text{bare}}$, was evaluated as the weighted average of the slopes obtained by the linear fittings of the trajectories in the rotating states.

The probability of the rewinding activation, $p$, was evaluated by manually counting the number of the activation processes accompanied by transient rewinding with the magnitude sufficiently larger than the angular fluctuation. $p$ were obtained by dividing the number by the total number of the activation events $n$. Error bars in Fig. 4b represent $\sqrt{np(1-p)}/n$, where $\sqrt{np(1-p)}$ is the standard deviation of the binomial distribution.

## II. RELATION WITH ELEMENTARY REACTIONS

We here characterize our model in terms of the reaction scheme (Fig. S8). $F_1$ has three catalytic sites [11, 12], each of which catalyzes the reaction of one ATP per 360° rotation of $\gamma$-shaft [13, 14]. The elementary reactions take place sequentially, accompanied by the rotation of the $\gamma$-shaft. The relation of the elementary reactions with the angle of the $\gamma$-shaft has been well investigated in the case of ATP hydrolysis. When the angle where ATP binds to one of the catalytic sites is set at 0°, this ATP is cleaved at 200° [15, 16], ADP is released at 240° [14, 17], and $P_i$ is released at 320° [18] on average. These sequential elementary reactions occur at three catalytic sites with a phase shift of ±120° from each other.

In the inhibited state, the angle of the $\gamma$-shaft settles in 320° (equivalent with 80° and 200°) [19]. The lapse into the inhibited state is triggered by the release of $P_i$ at 200°, not at 320° [20]. In addition, the release of ADP promoted by the forced rotation to the ATP-hydrolytic direction can activate the inhibited $F_1$ [10]. These facts indicate that the elementary reactions associated with the activation through $L$ and $S$ are the re-binding of $P_i$ at 200° and the release of ADP at > 320°, respectively. The increase in [ADP] suppressed the net rotation rate (Fig. 2c). The effect was larger for $N_{ex} > 0$ than $N_{ex} < 0$, suggesting the suppression of the activation through $S$. On the other hand, the decrease in [$P_i$] suppressed the net rotation rate for $N_{ex} < 0$ (Figs. S2, S4, and S9), suggesting the suppression of the activation through $L$. These tendencies are consistent with the expectations based on the reaction scheme.

## III. CONSISTENCY WITH THE PREVIOUS RESULTS

Hirono-Hara $et\ al.$ showed that the forced rotation activates a single $F_1$ molecule from the inhibited state by using magnetic tweezer [10]. They trapped the $\gamma$-shaft's angle of the inhibited $F_1$ at a certain angle for a finite duration and measured the activation probability. The activation was enhanced with the shift of the stall angle in the ATP-hydrolytic direction and not with that in the ATP-synthetic direction. Their result seems to be inconsistent with the fact that the external torque in the ATP-synthetic direction activates $F_1$. However, since they measured the probability of activation at the fixed angle, it saturates as the equilibration progresses between the rotating and inhibited states. Therefore, in their experiment, it is difficult to detect the activation at an angle where the inactivation frequently takes place, and we consider that this is the reason why they did not find the activation at $L$.

---

[1] Y. Rondelez, G. Tresset, T. Nakashima, Y. Kato-Yamada, H. Fujita, S. Takeuchi, and H. Noji, Nature **433**, 773 (2005).
[2] S. Toyabe and E. Muneyuki, New J. Phys. **17**, 015008 (2015).
[3] M. Washizu, Y. Kurahashi, H. Iochi, O. Kurosawa, S. Aizawa, S. Kudo, Y. Magariyama, and H. Hotani, IEEE Trans. Ind. Appl. **29**, 286 (1991).
[4] H. C. Berg and L. Turner, Biophys. J. **65**, 2201 (1993).
[5] T. Watanabe-Nakayama, S. Toyabe, S. Kudo, S. Sugiyama, M. Yoshida, and E. Muneyuki, Biochem. Biophys. Res. Comm. **366**, 951 (2008).
[6] S. Toyabe, T. Watanabe-Nakayama, T. Okamoto, S. Kudo, and E. Muneyuki, Proc. Nat. Acad. Sci. USA **108**, 17951 (2011).
[7] S. Toyabe, T. Okamoto, T. Watanabe-Nakayama, H. Taketani, S. Kudo, and E. Muneyuki, Phys. Rev. Lett. **104**, 198103 (2010).
[8] C. M. Bishop, *Pattern recognition and machine learning*, Information science and statistics (Springer, New York, 2006).
[9] H. Risken, *The Fokker-Planck equation: methods of solution and applications*, 2nd ed., Springer series in synergetics No. v. 18 (Springer-Verlag, New York, 1996).
[10] Y. Hirono-Hara, K. Ishizuka, K. Kinosita, Jr., M. Yoshida, and H. Noji, Proc. Natl. Acad. Sci. USA **102**, 4288 (2005).
[11] J. P. Abrahams, A. G. W. Leslie, R. Lutter, and J. E. Walker, Nature **370**, 621 (1994).
[12] Y. Shirakihara, A. Shiratori, H. Tanikawa, M. Nakasako, M. Yoshida, and T. Suzuki, FEBS J. **282**, 2895 (2015).
[13] R. Yasuda, H. Noji, K. Kinosita, Jr., and M. Yoshida, Cell **93**, 1117 (1998).
[14] T. Nishizaka, K. Oiwa, H. Noji, S. Kimura, E. Muneyuki, M. Yoshida, and K. Kinosita, Jr., Nat. Str. Mol. Biol **11**, 142 (2004).
[15] K. Shimabukuro, R. Yasuda, E. Muneyuki, K. Y. Hara, K. Kinosita, and M. Yoshida, Proc. Nat. Acad. Sci. **100**, 14731 (2003).
[16] T. Ariga, E. Muneyuki, and M. Yoshida, Nature Structural & Molecular Biology **14**, 841 (2007).
[17] K. Adachi, K. Oiwa, T. Nishizaka, S. Furuike, H. Noji, H. Itoh, M. Yoshida, and K. Kinosita, Jr., Cell **130**, 309 (2007).
[18] R. Watanabe, R. Iino, and H. Noji, Nat. Chem. Biol. **6**, 814 (2010).
[19] Y. Hirono-Hara, H. Noji, M. Nishiura, E. Muneyuki, K. Y. Hara, R. Yasuda, K. Kinosita, and M. Yoshida, Proc. Nat. Acad. Sci. **98**, 13649 (2001).

[20] R. Watanabe and H. Noji, Nature Comm. **5** (2014).

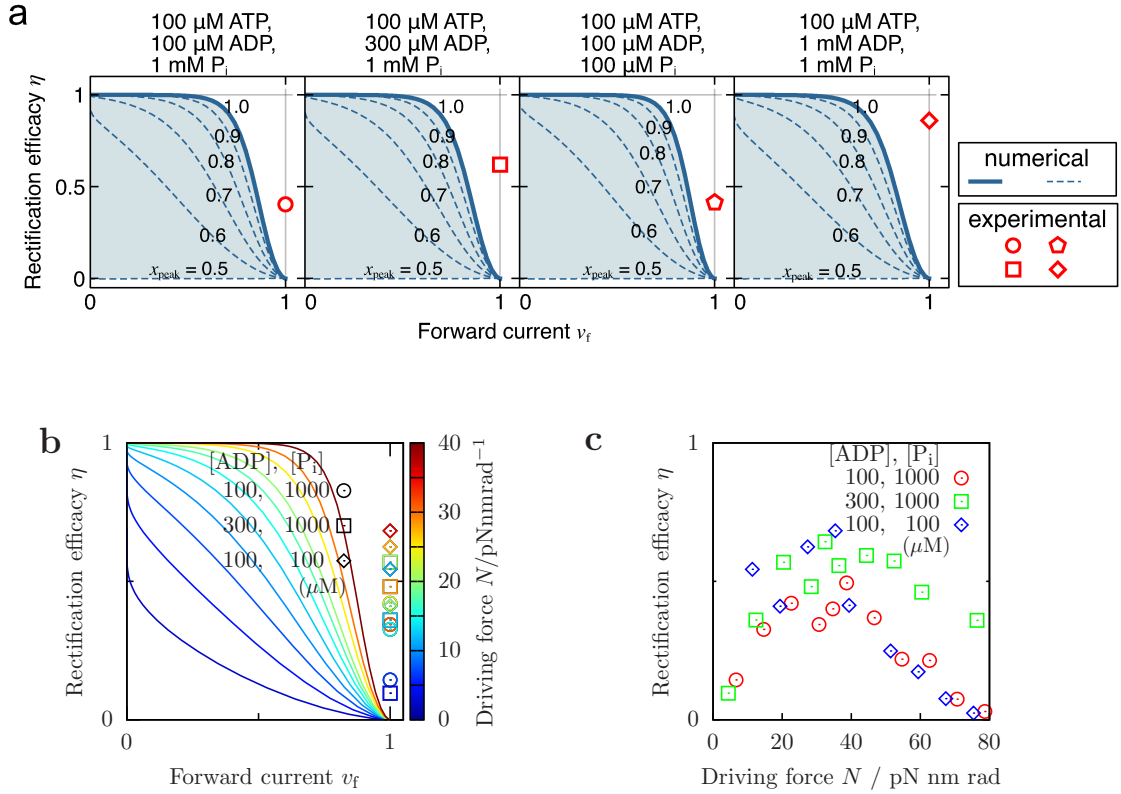

FIG. S1. The trade-off between the current and the rectification efficacy. **a.** The trade-off for each experimental condition. The values of the driving force used to calculate the trade-off for the simple ratchet model were set to $\Delta\mu$ per period at the indicated experimental conditions. In Fig. 1b, the trade-off only for one of the experimental conditions, [ATP] = [ADP] = 100 μM and [$P_i$] = 1 mM, is shown. As explained in Fig. 1b, we assumed $\tilde{v}_f = 1$ for the case of [ATP] = 100 μM, [ADP] = 1 mM, and [$P_i$] = 1 mM. **b.** The dependence of the trade-off on the driving force, $N$. The lines show the trade-off of the simple ratchet model with $x_{peak} = 1.0$. The symbols represent the experimental results of $F_1$. The colors of the lines and the symbols indicate the values of $N$. **c.** The dependence of $\eta$ on $N$ for each experimental condition.

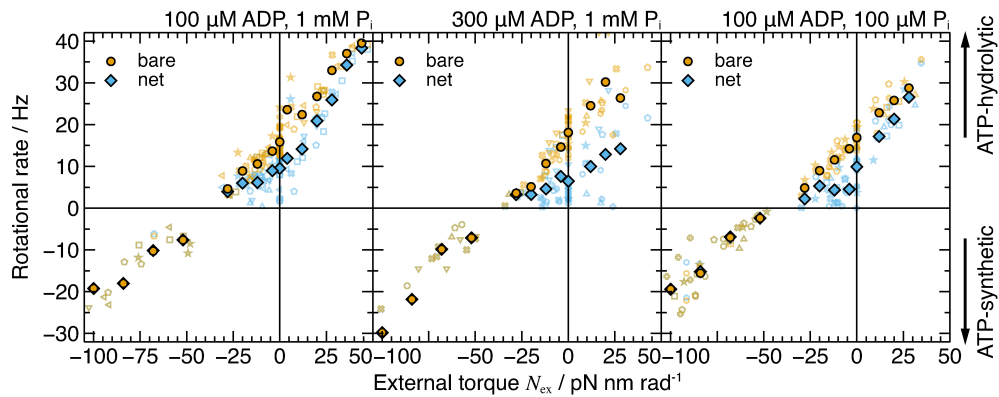

FIG. S2. Torque dependence of rotation rates for three conditions. The concentrations of ADP and $P_i$ are indicated in the figure. [ATP] = 100 μM. The results of each molecule are shown as light color symbols. The widths of bins are the same as in Fig. 2c.

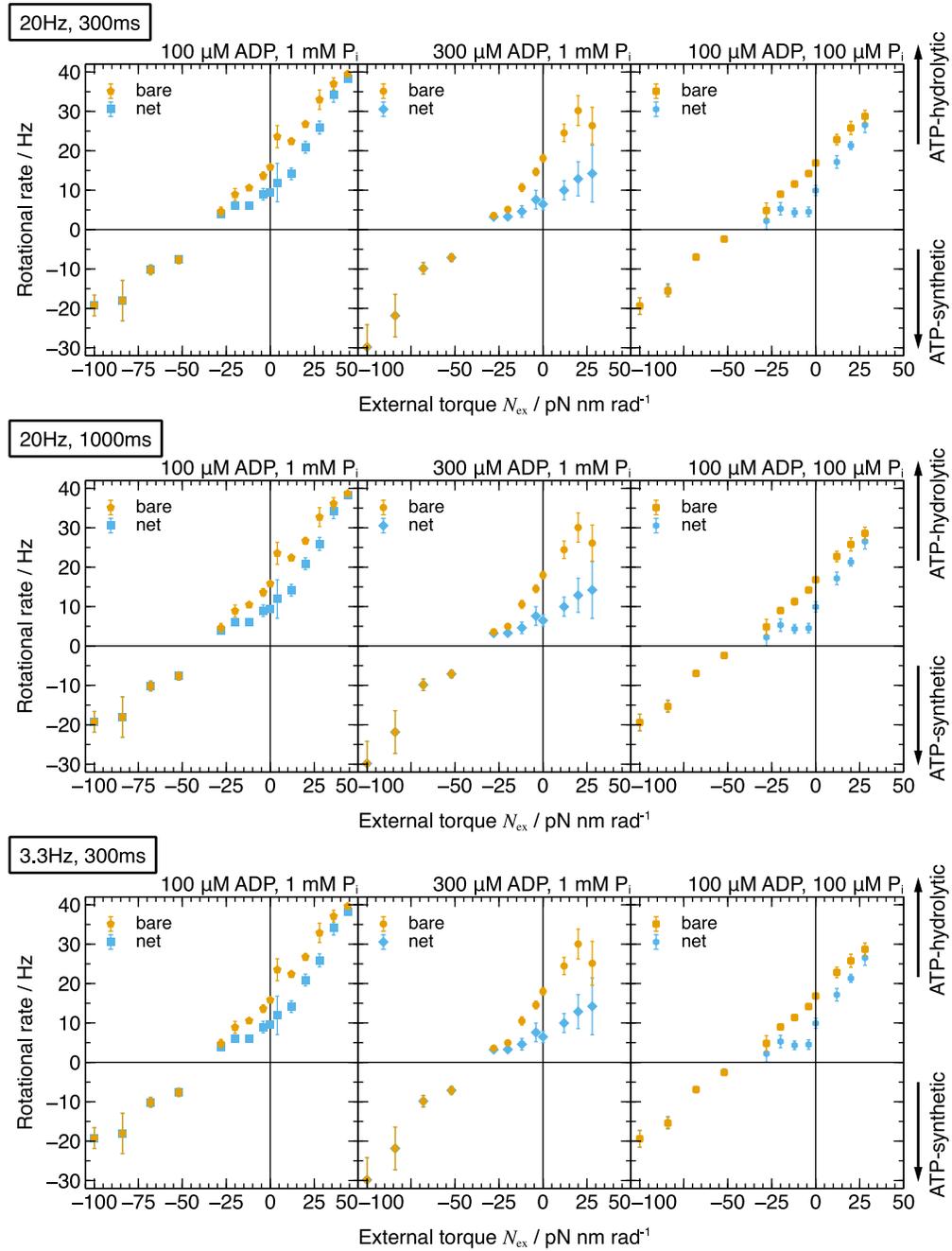

FIG. S3. The effect of the interval between selected frames and the threshold for the short pauses on the torque dependence of rotation rates. The interval between selected frames and the threshold for the short pauses were set to the values shown in each panel. The concentrations of ADP and $P_i$ are indicated in the figure. [ATP] = 100 μM. The widths of bins are the same as in Fig. 2c. Error bars denote standard errors of the mean. The differences are hardly seen among the different values of the interval between selected frames and the threshold for the short pauses.

8

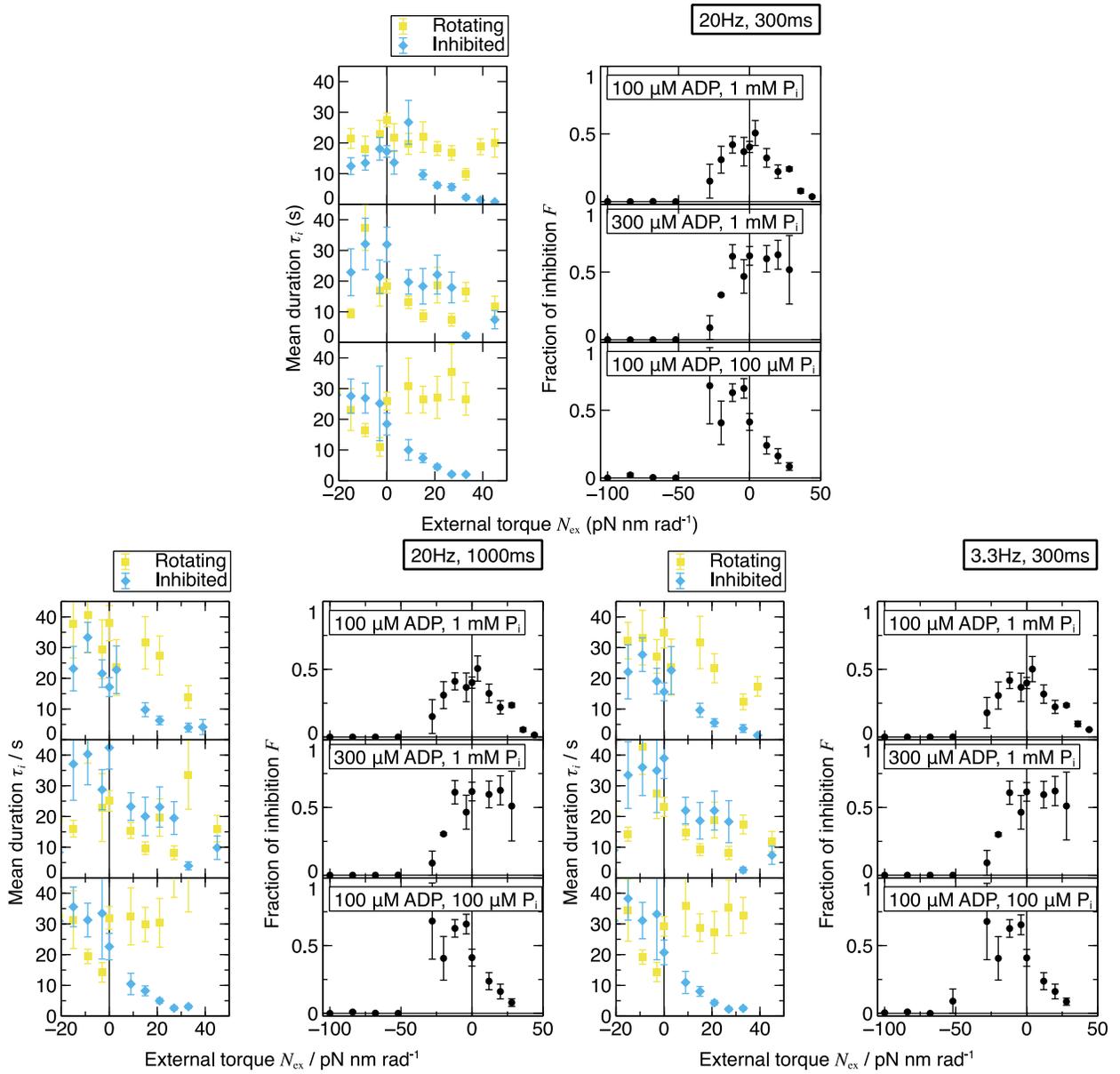

FIG. S4. The effect of the interval between selected frames and the threshold for the short pauses on the torque dependence of the mean duration $\tau_{\mathrm{rot}}$ and $\tau_{\mathrm{inh}}$, and the fraction of the time in the inhibited state $F$. The interval between selected frames and the threshold for the short pauses were set to the values shown in each panel. The concentrations of ADP and $P_i$ are indicated in the figure. [ATP] = 100 μM. The widths of bins are the same as in Fig. 3. Error bars are standard errors of the mean. The differences among the different values of the interval between selected frames and the threshold for the short pauses are just comparable with the experimental errors of the results.



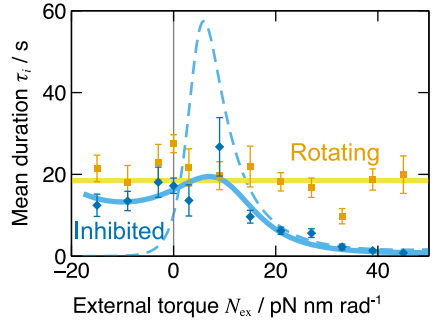

FIG. S7. The same as Fig. 3a in the main text, but the fitting curve obtained by fixing $\theta^* - L = 120°$ is plotted together as the blue dashed curve.

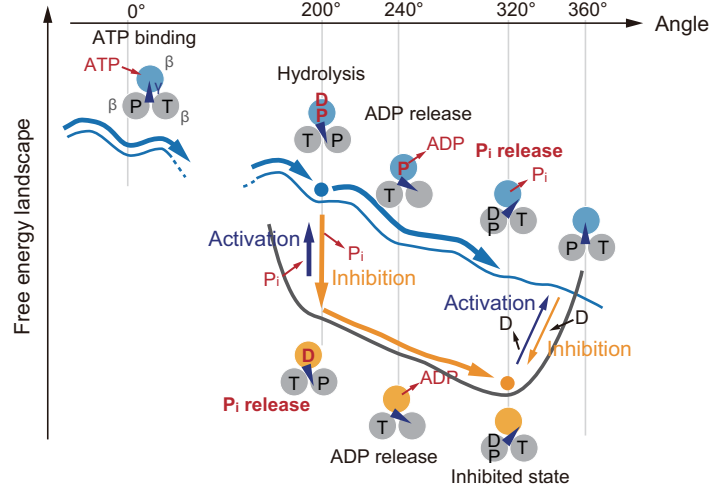

FIG. S8. Reaction scheme obtained previously [18] is superposed on the schematic of the free energy landscape. T, D, and P represent ATP, ADP, and $P_i$, respectively.

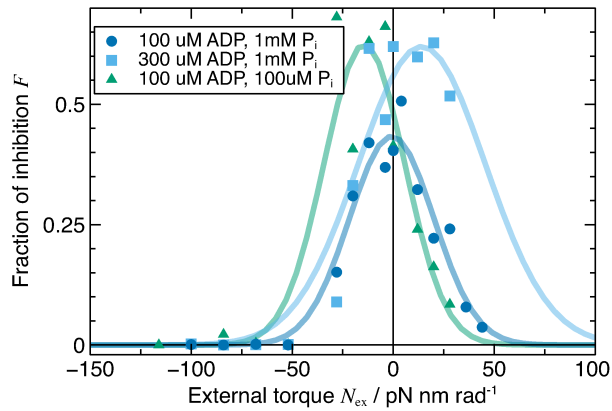

FIG. S9. The fraction of the time in the inhibited state for three conditions. [ATP] = 100 μM. The widths of bins are the same as in Fig. 3. The curves show the fitting by Gaussian functions. The increase in [ADP] suppressed the net rotation rate for $N_{\mathrm{ex}} > 0$ more, whereas the decrease in [$P_i$] suppressed the net rotation rate for $N_{\mathrm{ex}} < 0$ more.




\begin{thebibliography}{28}%
\makeatletter
\providecommand \@ifxundefined [1]{%
 \@ifx{#1\undefined}
}%
\providecommand \@ifnum [1]{%
 \ifnum #1\expandafter \@firstoftwo
 \else \expandafter \@secondoftwo
 \fi
}%
\providecommand \@ifx [1]{%
 \ifx #1\expandafter \@firstoftwo
 \else \expandafter \@secondoftwo
 \fi
}%
\providecommand \natexlab [1]{#1}%
\providecommand \enquote  [1]{``#1''}%
\providecommand \bibnamefont  [1]{#1}%
\providecommand \bibfnamefont [1]{#1}%
\providecommand \citenamefont [1]{#1}%
\providecommand \href@noop [0]{\@secondoftwo}%
\providecommand \href [0]{\begingroup \@sanitize@url \@href}%
\providecommand \@href[1]{\@@startlink{#1}\@@href}%
\providecommand \@@href[1]{\endgroup#1\@@endlink}%
\providecommand \@sanitize@url [0]{\catcode `\\12\catcode `\$12\catcode
  `\&12\catcode `\#12\catcode `\^12\catcode `\_12\catcode `\%12\relax}%
\providecommand \@@startlink[1]{}%
\providecommand \@@endlink[0]{}%
\providecommand \url  [0]{\begingroup\@sanitize@url \@url }%
\providecommand \@url [1]{\endgroup\@href {#1}{\urlprefix }}%
\providecommand \urlprefix  [0]{URL }%
\providecommand \Eprint [0]{\href }%
\providecommand \doibase [0]{https://doi.org/}%
\providecommand \selectlanguage [0]{\@gobble}%
\providecommand \bibinfo  [0]{\@secondoftwo}%
\providecommand \bibfield  [0]{\@secondoftwo}%
\providecommand \translation [1]{[#1]}%
\providecommand \BibitemOpen [0]{}%
\providecommand \bibitemStop [0]{}%
\providecommand \bibitemNoStop [0]{.\EOS\space}%
\providecommand \EOS [0]{\spacefactor3000\relax}%
\providecommand \BibitemShut  [1]{\csname bibitem#1\endcsname}%
\let\auto@bib@innerbib\@empty
\bibitem [{\citenamefont {Boyer}(1997)}]{Boyer1997}%
  \BibitemOpen
  \bibfield  {author} {\bibinfo {author} {\bibfnamefont {P.~D.}\ \bibnamefont
  {Boyer}},\ }\href@noop {} {\bibfield  {journal} {\bibinfo  {journal} {Annu.
  Rev. Biochem.}\ }\textbf {\bibinfo {volume} {66}},\ \bibinfo {pages} {717}
  (\bibinfo {year} {1997})}\BibitemShut {NoStop}%
\bibitem [{\citenamefont {Abrahams}\ \emph {et~al.}(1994)\citenamefont
  {Abrahams}, \citenamefont {Leslie}, \citenamefont {Lutter},\ and\
  \citenamefont {Walker}}]{Abrahams1994}%
  \BibitemOpen
  \bibfield  {author} {\bibinfo {author} {\bibfnamefont {J.~P.}\ \bibnamefont
  {Abrahams}}, \bibinfo {author} {\bibfnamefont {A.~G.~W.}\ \bibnamefont
  {Leslie}}, \bibinfo {author} {\bibfnamefont {R.}~\bibnamefont {Lutter}},\
  and\ \bibinfo {author} {\bibfnamefont {J.~E.}\ \bibnamefont {Walker}},\
  }\href@noop {} {\bibfield  {journal} {\bibinfo  {journal} {Nature}\ }\textbf
  {\bibinfo {volume} {370}},\ \bibinfo {pages} {621} (\bibinfo {year}
  {1994})}\BibitemShut {NoStop}%
\bibitem [{\citenamefont {Boyer}(1993)}]{boyer1993}%
  \BibitemOpen
  \bibfield  {author} {\bibinfo {author} {\bibfnamefont {P.~D.}\ \bibnamefont
  {Boyer}},\ }\href@noop {} {\bibfield  {journal} {\bibinfo  {journal}
  {Biochim. Biophys. Acta}\ }\textbf {\bibinfo {volume} {1140}},\ \bibinfo
  {pages} {215} (\bibinfo {year} {1993})}\BibitemShut {NoStop}%
\bibitem [{\citenamefont {Guo}\ \emph {et~al.}(2019)\citenamefont {Guo},
  \citenamefont {Suzuki},\ and\ \citenamefont {Rubinstein}}]{Guo2019}%
  \BibitemOpen
  \bibfield  {author} {\bibinfo {author} {\bibfnamefont {H.}~\bibnamefont
  {Guo}}, \bibinfo {author} {\bibfnamefont {T.}~\bibnamefont {Suzuki}},\ and\
  \bibinfo {author} {\bibfnamefont {J.~L.}\ \bibnamefont {Rubinstein}},\
  }\href@noop {} {\bibfield  {journal} {\bibinfo  {journal} {{eLife}}\ }\textbf
  {\bibinfo {volume} {8}},\ \bibinfo {pages} {e43128} (\bibinfo {year}
  {2019})}\BibitemShut {NoStop}%
\bibitem [{\citenamefont {Itoh}\ \emph {et~al.}(2004)\citenamefont {Itoh},
  \citenamefont {Takahashi}, \citenamefont {Adachi}, \citenamefont {Noji},
  \citenamefont {Yasuda}, \citenamefont {Yoshida},\ and\ \citenamefont
  {Kinosita}}]{Itoh2004}%
  \BibitemOpen
  \bibfield  {author} {\bibinfo {author} {\bibfnamefont {H.}~\bibnamefont
  {Itoh}}, \bibinfo {author} {\bibfnamefont {A.}~\bibnamefont {Takahashi}},
  \bibinfo {author} {\bibfnamefont {K.}~\bibnamefont {Adachi}}, \bibinfo
  {author} {\bibfnamefont {H.}~\bibnamefont {Noji}}, \bibinfo {author}
  {\bibfnamefont {R.}~\bibnamefont {Yasuda}}, \bibinfo {author} {\bibfnamefont
  {M.}~\bibnamefont {Yoshida}},\ and\ \bibinfo {author} {\bibfnamefont
  {K.}~\bibnamefont {Kinosita}, \bibfnamefont {Jr.}},\ }\href@noop {}
  {\bibfield  {journal} {\bibinfo  {journal} {Nature}\ }\textbf {\bibinfo
  {volume} {427}},\ \bibinfo {pages} {465} (\bibinfo {year}
  {2004})}\BibitemShut {NoStop}%
\bibitem [{\citenamefont {Rondelez}\ \emph {et~al.}(2005)\citenamefont
  {Rondelez}, \citenamefont {Tresset}, \citenamefont {Nakashima}, \citenamefont
  {Kato-Yamada}, \citenamefont {Fujita}, \citenamefont {Takeuchi},\ and\
  \citenamefont {Noji}}]{Rondelez2005}%
  \BibitemOpen
  \bibfield  {author} {\bibinfo {author} {\bibfnamefont {Y.}~\bibnamefont
  {Rondelez}}, \bibinfo {author} {\bibfnamefont {G.}~\bibnamefont {Tresset}},
  \bibinfo {author} {\bibfnamefont {T.}~\bibnamefont {Nakashima}}, \bibinfo
  {author} {\bibfnamefont {Y.}~\bibnamefont {Kato-Yamada}}, \bibinfo {author}
  {\bibfnamefont {H.}~\bibnamefont {Fujita}}, \bibinfo {author} {\bibfnamefont
  {S.}~\bibnamefont {Takeuchi}},\ and\ \bibinfo {author} {\bibfnamefont
  {H.}~\bibnamefont {Noji}},\ }\href@noop {} {\bibfield  {journal} {\bibinfo
  {journal} {Nature}\ }\textbf {\bibinfo {volume} {433}},\ \bibinfo {pages}
  {773} (\bibinfo {year} {2005})}\BibitemShut {NoStop}%
\bibitem [{\citenamefont {Toyabe}\ \emph {et~al.}(2011)\citenamefont {Toyabe},
  \citenamefont {Watanabe-Nakayama}, \citenamefont {Okamoto}, \citenamefont
  {Kudo},\ and\ \citenamefont {Muneyuki}}]{Toyabe2011}%
  \BibitemOpen
  \bibfield  {author} {\bibinfo {author} {\bibfnamefont {S.}~\bibnamefont
  {Toyabe}}, \bibinfo {author} {\bibfnamefont {T.}~\bibnamefont
  {Watanabe-Nakayama}}, \bibinfo {author} {\bibfnamefont {T.}~\bibnamefont
  {Okamoto}}, \bibinfo {author} {\bibfnamefont {S.}~\bibnamefont {Kudo}},\ and\
  \bibinfo {author} {\bibfnamefont {E.}~\bibnamefont {Muneyuki}},\ }\href@noop
  {} {\bibfield  {journal} {\bibinfo  {journal} {Proc. Nat. Acad. Sci. USA}\
  }\textbf {\bibinfo {volume} {108}},\ \bibinfo {pages} {17951} (\bibinfo
  {year} {2011})}\BibitemShut {NoStop}%
\bibitem [{\citenamefont {Saita}\ \emph {et~al.}(2015)\citenamefont {Saita},
  \citenamefont {Suzuki}, \citenamefont {Kinosita},\ and\ \citenamefont
  {Yoshida}}]{Saita2015}%
  \BibitemOpen
  \bibfield  {author} {\bibinfo {author} {\bibfnamefont {E.}~\bibnamefont
  {Saita}}, \bibinfo {author} {\bibfnamefont {T.}~\bibnamefont {Suzuki}},
  \bibinfo {author} {\bibfnamefont {K.}~\bibnamefont {Kinosita}},\ and\
  \bibinfo {author} {\bibfnamefont {M.}~\bibnamefont {Yoshida}},\ }\href@noop
  {} {\bibfield  {journal} {\bibinfo  {journal} {Proc. Nat. Acad. Sci.}\
  }\textbf {\bibinfo {volume} {112}},\ \bibinfo {pages} {9626} (\bibinfo {year}
  {2015})}\BibitemShut {NoStop}%
\bibitem [{\citenamefont {Soga}\ \emph {et~al.}(2017)\citenamefont {Soga},
  \citenamefont {Kimura}, \citenamefont {Kinosita}, \citenamefont {Yoshida},\
  and\ \citenamefont {Suzuki}}]{Soga2017}%
  \BibitemOpen
  \bibfield  {author} {\bibinfo {author} {\bibfnamefont {N.}~\bibnamefont
  {Soga}}, \bibinfo {author} {\bibfnamefont {K.}~\bibnamefont {Kimura}},
  \bibinfo {author} {\bibfnamefont {K.}~\bibnamefont {Kinosita}}, \bibinfo
  {author} {\bibfnamefont {M.}~\bibnamefont {Yoshida}},\ and\ \bibinfo {author}
  {\bibfnamefont {T.}~\bibnamefont {Suzuki}},\ }\href@noop {} {\bibfield
  {journal} {\bibinfo  {journal} {Proc. Nat. Acad. Sci.}\ }\textbf {\bibinfo
  {volume} {114}},\ \bibinfo {pages} {4960} (\bibinfo {year}
  {2017})}\BibitemShut {NoStop}%
\bibitem [{Note1()}]{Note1}%
  \BibitemOpen
  \bibinfo {note} {See the Supplemental Material at [URL] for additional text
  and Figs~S1--S9, which includes Refs~[29-40].}\BibitemShut {Stop}%
\bibitem [{\citenamefont {Bustamante}\ \emph {et~al.}(2005)\citenamefont
  {Bustamante}, \citenamefont {Liphardt},\ and\ \citenamefont
  {Ritort}}]{Bustamante2005}%
  \BibitemOpen
  \bibfield  {author} {\bibinfo {author} {\bibfnamefont {C.}~\bibnamefont
  {Bustamante}}, \bibinfo {author} {\bibfnamefont {J.}~\bibnamefont
  {Liphardt}},\ and\ \bibinfo {author} {\bibfnamefont {F.}~\bibnamefont
  {Ritort}},\ }\href@noop {} {\bibfield  {journal} {\bibinfo  {journal}
  {Physics Today}\ }\textbf {\bibinfo {volume} {58}},\ \bibinfo {pages} {43}
  (\bibinfo {year} {2005})}\BibitemShut {NoStop}%
\bibitem [{\citenamefont {Astumian}(1997)}]{astumian1997}%
  \BibitemOpen
  \bibfield  {author} {\bibinfo {author} {\bibfnamefont {R.~D.}\ \bibnamefont
  {Astumian}},\ }\href {https://doi.org/10.1126/science.276.5314.917}
  {\bibfield  {journal} {\bibinfo  {journal} {Science}\ }\textbf {\bibinfo
  {volume} {276}},\ \bibinfo {pages} {917} (\bibinfo {year}
  {1997})}\BibitemShut {NoStop}%
\bibitem [{\citenamefont {Reimann}(2002)}]{Reimann2002}%
  \BibitemOpen
  \bibfield  {author} {\bibinfo {author} {\bibfnamefont {P.}~\bibnamefont
  {Reimann}},\ }\href@noop {} {\bibfield  {journal} {\bibinfo  {journal} {Phys.
  Rep.}\ }\textbf {\bibinfo {volume} {361}},\ \bibinfo {pages} {57} (\bibinfo
  {year} {2002})}\BibitemShut {NoStop}%
\bibitem [{\citenamefont {Syroeshkin}\ \emph {et~al.}(1995)\citenamefont
  {Syroeshkin}, \citenamefont {Vasilyeva},\ and\ \citenamefont
  {Vinogradov}}]{syroeshkin-atp-1995}%
  \BibitemOpen
  \bibfield  {author} {\bibinfo {author} {\bibfnamefont {A.}~\bibnamefont
  {Syroeshkin}}, \bibinfo {author} {\bibfnamefont {E.}~\bibnamefont
  {Vasilyeva}},\ and\ \bibinfo {author} {\bibfnamefont {A.}~\bibnamefont
  {Vinogradov}},\ }\href@noop {} {\bibfield  {journal} {\bibinfo  {journal}
  {FEBS Lett.}\ }\textbf {\bibinfo {volume} {366}},\ \bibinfo {pages} {29}
  (\bibinfo {year} {1995})}\BibitemShut {NoStop}%
\bibitem [{\citenamefont {Bald}\ \emph {et~al.}(1998)\citenamefont {Bald},
  \citenamefont {Amano}, \citenamefont {Muneyuki}, \citenamefont {Pitard},
  \citenamefont {Rigaud}, \citenamefont {Kruip}, \citenamefont {Hisabori},
  \citenamefont {Yoshida},\ and\ \citenamefont {Shibata}}]{bald-atp-1998}%
  \BibitemOpen
  \bibfield  {author} {\bibinfo {author} {\bibfnamefont {D.}~\bibnamefont
  {Bald}}, \bibinfo {author} {\bibfnamefont {T.}~\bibnamefont {Amano}},
  \bibinfo {author} {\bibfnamefont {E.}~\bibnamefont {Muneyuki}}, \bibinfo
  {author} {\bibfnamefont {B.}~\bibnamefont {Pitard}}, \bibinfo {author}
  {\bibfnamefont {J.-L.}\ \bibnamefont {Rigaud}}, \bibinfo {author}
  {\bibfnamefont {J.}~\bibnamefont {Kruip}}, \bibinfo {author} {\bibfnamefont
  {T.}~\bibnamefont {Hisabori}}, \bibinfo {author} {\bibfnamefont
  {M.}~\bibnamefont {Yoshida}},\ and\ \bibinfo {author} {\bibfnamefont
  {M.}~\bibnamefont {Shibata}},\ }\href@noop {} {\bibfield  {journal} {\bibinfo
   {journal} {J. Biol. Chem.}\ }\textbf {\bibinfo {volume} {273}},\ \bibinfo
  {pages} {865} (\bibinfo {year} {1998})}\BibitemShut {NoStop}%
\bibitem [{\citenamefont {Galkin}\ and\ \citenamefont
  {Vinogradov}(1999)}]{galkin-energy-dependent-1999}%
  \BibitemOpen
  \bibfield  {author} {\bibinfo {author} {\bibfnamefont {M.}~\bibnamefont
  {Galkin}}\ and\ \bibinfo {author} {\bibfnamefont {A.}~\bibnamefont
  {Vinogradov}},\ }\href@noop {} {\bibfield  {journal} {\bibinfo  {journal}
  {FEBS Lett.}\ }\textbf {\bibinfo {volume} {448}},\ \bibinfo {pages} {123}
  (\bibinfo {year} {1999})}\BibitemShut {NoStop}%
\bibitem [{\citenamefont {Noji}\ \emph {et~al.}(1997)\citenamefont {Noji},
  \citenamefont {Yasuda}, \citenamefont {Yoshida},\ and\ \citenamefont
  {Kinosita}}]{Noji1997}%
  \BibitemOpen
  \bibfield  {author} {\bibinfo {author} {\bibfnamefont {H.}~\bibnamefont
  {Noji}}, \bibinfo {author} {\bibfnamefont {R.}~\bibnamefont {Yasuda}},
  \bibinfo {author} {\bibfnamefont {M.}~\bibnamefont {Yoshida}},\ and\ \bibinfo
  {author} {\bibfnamefont {K.}~\bibnamefont {Kinosita}},\ }\href@noop {}
  {\bibfield  {journal} {\bibinfo  {journal} {Nature}\ }\textbf {\bibinfo
  {volume} {386}},\ \bibinfo {pages} {299} (\bibinfo {year}
  {1997})}\BibitemShut {NoStop}%
\bibitem [{\citenamefont {Hirono-Hara}\ \emph {et~al.}(2001)\citenamefont
  {Hirono-Hara}, \citenamefont {Noji}, \citenamefont {Nishiura}, \citenamefont
  {Muneyuki}, \citenamefont {Hara}, \citenamefont {Yasuda}, \citenamefont
  {Kinosita},\ and\ \citenamefont {Yoshida}}]{HironoHara2001}%
  \BibitemOpen
  \bibfield  {author} {\bibinfo {author} {\bibfnamefont {Y.}~\bibnamefont
  {Hirono-Hara}}, \bibinfo {author} {\bibfnamefont {H.}~\bibnamefont {Noji}},
  \bibinfo {author} {\bibfnamefont {M.}~\bibnamefont {Nishiura}}, \bibinfo
  {author} {\bibfnamefont {E.}~\bibnamefont {Muneyuki}}, \bibinfo {author}
  {\bibfnamefont {K.~Y.}\ \bibnamefont {Hara}}, \bibinfo {author}
  {\bibfnamefont {R.}~\bibnamefont {Yasuda}}, \bibinfo {author} {\bibfnamefont
  {K.}~\bibnamefont {Kinosita}},\ and\ \bibinfo {author} {\bibfnamefont
  {M.}~\bibnamefont {Yoshida}},\ }\href@noop {} {\bibfield  {journal} {\bibinfo
   {journal} {Proc. Nat. Acad. Sci.}\ }\textbf {\bibinfo {volume} {98}},\
  \bibinfo {pages} {13649} (\bibinfo {year} {2001})}\BibitemShut {NoStop}%
\bibitem [{\citenamefont {Hirono-Hara}\ \emph {et~al.}(2005)\citenamefont
  {Hirono-Hara}, \citenamefont {Ishizuka}, \citenamefont {Kinosita},
  \citenamefont {Yoshida},\ and\ \citenamefont {Noji}}]{Hirono-Hara2005}%
  \BibitemOpen
  \bibfield  {author} {\bibinfo {author} {\bibfnamefont {Y.}~\bibnamefont
  {Hirono-Hara}}, \bibinfo {author} {\bibfnamefont {K.}~\bibnamefont
  {Ishizuka}}, \bibinfo {author} {\bibfnamefont {K.}~\bibnamefont {Kinosita},
  \bibfnamefont {Jr.}}, \bibinfo {author} {\bibfnamefont {M.}~\bibnamefont
  {Yoshida}},\ and\ \bibinfo {author} {\bibfnamefont {H.}~\bibnamefont
  {Noji}},\ }\href@noop {} {\bibfield  {journal} {\bibinfo  {journal} {Proc.
  Natl. Acad. Sci. USA}\ }\textbf {\bibinfo {volume} {102}},\ \bibinfo {pages}
  {4288} (\bibinfo {year} {2005})}\BibitemShut {NoStop}%
\bibitem [{\citenamefont {Watanabe-Nakayama}\ \emph {et~al.}(2008)\citenamefont
  {Watanabe-Nakayama}, \citenamefont {Toyabe}, \citenamefont {Kudo},
  \citenamefont {Sugiyama}, \citenamefont {Yoshida},\ and\ \citenamefont
  {Muneyuki}}]{Watanabe-Nakayama2008}%
  \BibitemOpen
  \bibfield  {author} {\bibinfo {author} {\bibfnamefont {T.}~\bibnamefont
  {Watanabe-Nakayama}}, \bibinfo {author} {\bibfnamefont {S.}~\bibnamefont
  {Toyabe}}, \bibinfo {author} {\bibfnamefont {S.}~\bibnamefont {Kudo}},
  \bibinfo {author} {\bibfnamefont {S.}~\bibnamefont {Sugiyama}}, \bibinfo
  {author} {\bibfnamefont {M.}~\bibnamefont {Yoshida}},\ and\ \bibinfo {author}
  {\bibfnamefont {E.}~\bibnamefont {Muneyuki}},\ }\href@noop {} {\bibfield
  {journal} {\bibinfo  {journal} {Biochem. Biophys. Res. Comm.}\ }\textbf
  {\bibinfo {volume} {366}},\ \bibinfo {pages} {951} (\bibinfo {year}
  {2008})}\BibitemShut {NoStop}%
\bibitem [{\citenamefont {Toyabe}\ \emph {et~al.}(2010)\citenamefont {Toyabe},
  \citenamefont {Okamoto}, \citenamefont {Watanabe-Nakayama}, \citenamefont
  {Taketani}, \citenamefont {Kudo},\ and\ \citenamefont
  {Muneyuki}}]{Toyabe2010PRL}%
  \BibitemOpen
  \bibfield  {author} {\bibinfo {author} {\bibfnamefont {S.}~\bibnamefont
  {Toyabe}}, \bibinfo {author} {\bibfnamefont {T.}~\bibnamefont {Okamoto}},
  \bibinfo {author} {\bibfnamefont {T.}~\bibnamefont {Watanabe-Nakayama}},
  \bibinfo {author} {\bibfnamefont {H.}~\bibnamefont {Taketani}}, \bibinfo
  {author} {\bibfnamefont {S.}~\bibnamefont {Kudo}},\ and\ \bibinfo {author}
  {\bibfnamefont {E.}~\bibnamefont {Muneyuki}},\ }\href@noop {} {\bibfield
  {journal} {\bibinfo  {journal} {Phys. Rev. Lett.}\ }\textbf {\bibinfo
  {volume} {104}},\ \bibinfo {pages} {198103} (\bibinfo {year}
  {2010})}\BibitemShut {NoStop}%
\bibitem [{\citenamefont {Krab}\ and\ \citenamefont {{van
  Wezel}}(1992)}]{Krab1992}%
  \BibitemOpen
  \bibfield  {author} {\bibinfo {author} {\bibfnamefont {K.}~\bibnamefont
  {Krab}}\ and\ \bibinfo {author} {\bibfnamefont {J.}~\bibnamefont {{van
  Wezel}}},\ }\href@noop {} {\bibfield  {journal} {\bibinfo  {journal}
  {Biochim. Biophys. Acta}\ }\textbf {\bibinfo {volume} {1098}},\ \bibinfo
  {pages} {172} (\bibinfo {year} {1992})}\BibitemShut {NoStop}%
\bibitem [{\citenamefont {Watanabe}\ \emph {et~al.}(2012)\citenamefont
  {Watanabe}, \citenamefont {Okuno}, \citenamefont {Sakakihara}, \citenamefont
  {Shimabukuro}, \citenamefont {Iino}, \citenamefont {Yoshida},\ and\
  \citenamefont {Noji}}]{watanabe-mechanical-2012}%
  \BibitemOpen
  \bibfield  {author} {\bibinfo {author} {\bibfnamefont {R.}~\bibnamefont
  {Watanabe}}, \bibinfo {author} {\bibfnamefont {D.}~\bibnamefont {Okuno}},
  \bibinfo {author} {\bibfnamefont {S.}~\bibnamefont {Sakakihara}}, \bibinfo
  {author} {\bibfnamefont {K.}~\bibnamefont {Shimabukuro}}, \bibinfo {author}
  {\bibfnamefont {R.}~\bibnamefont {Iino}}, \bibinfo {author} {\bibfnamefont
  {M.}~\bibnamefont {Yoshida}},\ and\ \bibinfo {author} {\bibfnamefont
  {H.}~\bibnamefont {Noji}},\ }\href@noop {} {\bibfield  {journal} {\bibinfo
  {journal} {Nat. Chem. Biol.}\ }\textbf {\bibinfo {volume} {8}},\ \bibinfo
  {pages} {86} (\bibinfo {year} {2012})}\BibitemShut {NoStop}%
\bibitem [{\citenamefont {Watanabe}\ \emph {et~al.}(2010)\citenamefont
  {Watanabe}, \citenamefont {Iino},\ and\ \citenamefont {Noji}}]{Watanabe2010}%
  \BibitemOpen
  \bibfield  {author} {\bibinfo {author} {\bibfnamefont {R.}~\bibnamefont
  {Watanabe}}, \bibinfo {author} {\bibfnamefont {R.}~\bibnamefont {Iino}},\
  and\ \bibinfo {author} {\bibfnamefont {H.}~\bibnamefont {Noji}},\ }\href@noop
  {} {\bibfield  {journal} {\bibinfo  {journal} {Nat. Chem. Biol.}\ }\textbf
  {\bibinfo {volume} {6}},\ \bibinfo {pages} {814} (\bibinfo {year}
  {2010})}\BibitemShut {NoStop}%
\bibitem [{\citenamefont {Seifert}(2012)}]{Seifert2012}%
  \BibitemOpen
  \bibfield  {author} {\bibinfo {author} {\bibfnamefont {U.}~\bibnamefont
  {Seifert}},\ }\href@noop {} {\bibfield  {journal} {\bibinfo  {journal} {Rep.
  Prog. Phys.}\ }\textbf {\bibinfo {volume} {75}},\ \bibinfo {pages} {126001}
  (\bibinfo {year} {2012})}\BibitemShut {NoStop}%
\bibitem [{\citenamefont {Barato}\ and\ \citenamefont
  {Seifert}(2015)}]{barato2015}%
  \BibitemOpen
  \bibfield  {author} {\bibinfo {author} {\bibfnamefont {A.~C.}\ \bibnamefont
  {Barato}}\ and\ \bibinfo {author} {\bibfnamefont {U.}~\bibnamefont
  {Seifert}},\ }\href {https://doi.org/10.1103/PhysRevLett.114.158101}
  {\bibfield  {journal} {\bibinfo  {journal} {Phys. Rev. Lett.}\ }\textbf
  {\bibinfo {volume} {114}},\ \bibinfo {pages} {158101} (\bibinfo {year}
  {2015})}\BibitemShut {NoStop}%
\bibitem [{\citenamefont {Dechant}\ and\ \citenamefont
  {Sasa}(2020)}]{dechant2020}%
  \BibitemOpen
  \bibfield  {author} {\bibinfo {author} {\bibfnamefont {A.}~\bibnamefont
  {Dechant}}\ and\ \bibinfo {author} {\bibfnamefont {S.-i.}\ \bibnamefont
  {Sasa}},\ }\href {https://doi.org/10.1073/pnas.1918386117} {\bibfield
  {journal} {\bibinfo  {journal} {Proc. Natl. Acad. Sci. USA}\ }\textbf
  {\bibinfo {volume} {117}},\ \bibinfo {pages} {6430} (\bibinfo {year}
  {2020})}\BibitemShut {NoStop}%
\bibitem [{\citenamefont {Toyabe}\ \emph {et~al.}(2012)\citenamefont {Toyabe},
  \citenamefont {Ueno},\ and\ \citenamefont {Muneyuki}}]{Toyabe2012}%
  \BibitemOpen
  \bibfield  {author} {\bibinfo {author} {\bibfnamefont {S.}~\bibnamefont
  {Toyabe}}, \bibinfo {author} {\bibfnamefont {H.}~\bibnamefont {Ueno}},\ and\
  \bibinfo {author} {\bibfnamefont {E.}~\bibnamefont {Muneyuki}},\ }\href@noop
  {} {\bibfield  {journal} {\bibinfo  {journal} {EPL}\ }\textbf {\bibinfo
  {volume} {97}},\ \bibinfo {pages} {40004} (\bibinfo {year}
  {2012})}\BibitemShut {NoStop}%
\bibitem [{\citenamefont {Toyabe}\ and\ \citenamefont
  {Muneyuki}(2015)}]{Toyabe2015}%
  \BibitemOpen
  \bibfield  {author} {\bibinfo {author} {\bibfnamefont {S.}~\bibnamefont
  {Toyabe}}\ and\ \bibinfo {author} {\bibfnamefont {E.}~\bibnamefont
  {Muneyuki}},\ }\href@noop {} {\bibfield  {journal} {\bibinfo  {journal} {New
  J. Phys.}\ }\textbf {\bibinfo {volume} {17}},\ \bibinfo {pages} {015008}
  (\bibinfo {year} {2015})}\BibitemShut {NoStop}%
\bibitem [{\citenamefont {Washizu}\ \emph {et~al.}(1991)\citenamefont
  {Washizu}, \citenamefont {Kurahashi}, \citenamefont {Iochi}, \citenamefont
  {Kurosawa}, \citenamefont {Aizawa}, \citenamefont {Kudo}, \citenamefont
  {Magariyama},\ and\ \citenamefont {Hotani}}]{Washizu1991}%
  \BibitemOpen
  \bibfield  {author} {\bibinfo {author} {\bibfnamefont {M.}~\bibnamefont
  {Washizu}}, \bibinfo {author} {\bibfnamefont {Y.}~\bibnamefont {Kurahashi}},
  \bibinfo {author} {\bibfnamefont {H.}~\bibnamefont {Iochi}}, \bibinfo
  {author} {\bibfnamefont {O.}~\bibnamefont {Kurosawa}}, \bibinfo {author}
  {\bibfnamefont {S.}~\bibnamefont {Aizawa}}, \bibinfo {author} {\bibfnamefont
  {S.}~\bibnamefont {Kudo}}, \bibinfo {author} {\bibfnamefont {Y.}~\bibnamefont
  {Magariyama}},\ and\ \bibinfo {author} {\bibfnamefont {H.}~\bibnamefont
  {Hotani}},\ }\href@noop {} {\bibfield  {journal} {\bibinfo  {journal} {IEEE
  Trans. Ind. Appl.}\ }\textbf {\bibinfo {volume} {29}},\ \bibinfo {pages}
  {286} (\bibinfo {year} {1991})}\BibitemShut {NoStop}%
\bibitem [{\citenamefont {Berg}\ and\ \citenamefont {Turner}(1993)}]{Berg1993}%
  \BibitemOpen
  \bibfield  {author} {\bibinfo {author} {\bibfnamefont {H.~C.}\ \bibnamefont
  {Berg}}\ and\ \bibinfo {author} {\bibfnamefont {L.}~\bibnamefont {Turner}},\
  }\href@noop {} {\bibfield  {journal} {\bibinfo  {journal} {Biophys. J.}\
  }\textbf {\bibinfo {volume} {65}},\ \bibinfo {pages} {2201} (\bibinfo {year}
  {1993})}\BibitemShut {NoStop}%
\bibitem [{\citenamefont {Bishop}(2006)}]{bishop-pattern-2006}%
  \BibitemOpen
  \bibfield  {author} {\bibinfo {author} {\bibfnamefont {C.~M.}\ \bibnamefont
  {Bishop}},\ }\href@noop {} {\emph {\bibinfo {title} {Pattern recognition and
  machine learning}}},\ Information science and statistics\ (\bibinfo
  {publisher} {Springer},\ \bibinfo {address} {New York},\ \bibinfo {year}
  {2006})\BibitemShut {NoStop}%
\bibitem [{\citenamefont {Risken}(1996)}]{Risken}%
  \BibitemOpen
  \bibfield  {author} {\bibinfo {author} {\bibfnamefont {H.}~\bibnamefont
  {Risken}},\ }\href@noop {} {\emph {\bibinfo {title} {The {Fokker}-{Planck}
  equation: methods of solution and applications}}},\ \bibinfo {edition} {2nd}\
  ed.,\ \bibinfo {series} {Springer series in synergetics}\ No.\ \bibinfo
  {number} {v. 18}\ (\bibinfo  {publisher} {Springer-Verlag},\ \bibinfo
  {address} {New York},\ \bibinfo {year} {1996})\BibitemShut {NoStop}%
\bibitem [{\citenamefont {Shirakihara}\ \emph {et~al.}(2015)\citenamefont
  {Shirakihara}, \citenamefont {Shiratori}, \citenamefont {Tanikawa},
  \citenamefont {Nakasako}, \citenamefont {Yoshida},\ and\ \citenamefont
  {Suzuki}}]{shirakihara-structure-2015}%
  \BibitemOpen
  \bibfield  {author} {\bibinfo {author} {\bibfnamefont {Y.}~\bibnamefont
  {Shirakihara}}, \bibinfo {author} {\bibfnamefont {A.}~\bibnamefont
  {Shiratori}}, \bibinfo {author} {\bibfnamefont {H.}~\bibnamefont {Tanikawa}},
  \bibinfo {author} {\bibfnamefont {M.}~\bibnamefont {Nakasako}}, \bibinfo
  {author} {\bibfnamefont {M.}~\bibnamefont {Yoshida}},\ and\ \bibinfo {author}
  {\bibfnamefont {T.}~\bibnamefont {Suzuki}},\ }\href@noop {} {\bibfield
  {journal} {\bibinfo  {journal} {FEBS J.}\ }\textbf {\bibinfo {volume}
  {282}},\ \bibinfo {pages} {2895} (\bibinfo {year} {2015})}\BibitemShut
  {NoStop}%
\bibitem [{\citenamefont {Yasuda}\ \emph {et~al.}(1998)\citenamefont {Yasuda},
  \citenamefont {Noji}, \citenamefont {Kinosita},\ and\ \citenamefont
  {Yoshida}}]{Yasuda1998}%
  \BibitemOpen
  \bibfield  {author} {\bibinfo {author} {\bibfnamefont {R.}~\bibnamefont
  {Yasuda}}, \bibinfo {author} {\bibfnamefont {H.}~\bibnamefont {Noji}},
  \bibinfo {author} {\bibfnamefont {K.}~\bibnamefont {Kinosita}, \bibfnamefont
  {Jr.}},\ and\ \bibinfo {author} {\bibfnamefont {M.}~\bibnamefont {Yoshida}},\
  }\href@noop {} {\bibfield  {journal} {\bibinfo  {journal} {Cell}\ }\textbf
  {\bibinfo {volume} {93}},\ \bibinfo {pages} {1117} (\bibinfo {year}
  {1998})}\BibitemShut {NoStop}%
\bibitem [{\citenamefont {Nishizaka}\ \emph {et~al.}(2004)\citenamefont
  {Nishizaka}, \citenamefont {Oiwa}, \citenamefont {Noji}, \citenamefont
  {Kimura}, \citenamefont {Muneyuki}, \citenamefont {Yoshida},\ and\
  \citenamefont {Kinosita}}]{Nishizaka2004}%
  \BibitemOpen
  \bibfield  {author} {\bibinfo {author} {\bibfnamefont {T.}~\bibnamefont
  {Nishizaka}}, \bibinfo {author} {\bibfnamefont {K.}~\bibnamefont {Oiwa}},
  \bibinfo {author} {\bibfnamefont {H.}~\bibnamefont {Noji}}, \bibinfo {author}
  {\bibfnamefont {S.}~\bibnamefont {Kimura}}, \bibinfo {author} {\bibfnamefont
  {E.}~\bibnamefont {Muneyuki}}, \bibinfo {author} {\bibfnamefont
  {M.}~\bibnamefont {Yoshida}},\ and\ \bibinfo {author} {\bibfnamefont
  {K.}~\bibnamefont {Kinosita}, \bibfnamefont {Jr.}},\ }\href@noop {}
  {\bibfield  {journal} {\bibinfo  {journal} {Nat. Str. Mol. Biol}\ }\textbf
  {\bibinfo {volume} {11}},\ \bibinfo {pages} {142} (\bibinfo {year}
  {2004})}\BibitemShut {NoStop}%
\bibitem [{\citenamefont {Shimabukuro}\ \emph {et~al.}(2003)\citenamefont
  {Shimabukuro}, \citenamefont {Yasuda}, \citenamefont {Muneyuki},
  \citenamefont {Hara}, \citenamefont {Kinosita},\ and\ \citenamefont
  {Yoshida}}]{Shimabukuro2003}%
  \BibitemOpen
  \bibfield  {author} {\bibinfo {author} {\bibfnamefont {K.}~\bibnamefont
  {Shimabukuro}}, \bibinfo {author} {\bibfnamefont {R.}~\bibnamefont {Yasuda}},
  \bibinfo {author} {\bibfnamefont {E.}~\bibnamefont {Muneyuki}}, \bibinfo
  {author} {\bibfnamefont {K.~Y.}\ \bibnamefont {Hara}}, \bibinfo {author}
  {\bibfnamefont {K.}~\bibnamefont {Kinosita}},\ and\ \bibinfo {author}
  {\bibfnamefont {M.}~\bibnamefont {Yoshida}},\ }\href@noop {} {\bibfield
  {journal} {\bibinfo  {journal} {Proc. Nat. Acad. Sci.}\ }\textbf {\bibinfo
  {volume} {100}},\ \bibinfo {pages} {14731} (\bibinfo {year}
  {2003})}\BibitemShut {NoStop}%
\bibitem [{\citenamefont {Ariga}\ \emph {et~al.}(2007)\citenamefont {Ariga},
  \citenamefont {Muneyuki},\ and\ \citenamefont {Yoshida}}]{ariga2007}%
  \BibitemOpen
  \bibfield  {author} {\bibinfo {author} {\bibfnamefont {T.}~\bibnamefont
  {Ariga}}, \bibinfo {author} {\bibfnamefont {E.}~\bibnamefont {Muneyuki}},\
  and\ \bibinfo {author} {\bibfnamefont {M.}~\bibnamefont {Yoshida}},\ }\href
  {https://doi.org/http://dx.doi.org/10.1038/nsmb1296} {\bibfield  {journal}
  {\bibinfo  {journal} {Nature Structural \& Molecular Biology}\ }\textbf
  {\bibinfo {volume} {14}},\ \bibinfo {pages} {841} (\bibinfo {year}
  {2007})}\BibitemShut {NoStop}%
\bibitem [{\citenamefont {Adachi}\ \emph {et~al.}(2007)\citenamefont {Adachi},
  \citenamefont {Oiwa}, \citenamefont {Nishizaka}, \citenamefont {Furuike},
  \citenamefont {Noji}, \citenamefont {Itoh}, \citenamefont {Yoshida},\ and\
  \citenamefont {Kinosita}}]{Adachi2007}%
  \BibitemOpen
  \bibfield  {author} {\bibinfo {author} {\bibfnamefont {K.}~\bibnamefont
  {Adachi}}, \bibinfo {author} {\bibfnamefont {K.}~\bibnamefont {Oiwa}},
  \bibinfo {author} {\bibfnamefont {T.}~\bibnamefont {Nishizaka}}, \bibinfo
  {author} {\bibfnamefont {S.}~\bibnamefont {Furuike}}, \bibinfo {author}
  {\bibfnamefont {H.}~\bibnamefont {Noji}}, \bibinfo {author} {\bibfnamefont
  {H.}~\bibnamefont {Itoh}}, \bibinfo {author} {\bibfnamefont {M.}~\bibnamefont
  {Yoshida}},\ and\ \bibinfo {author} {\bibfnamefont {K.}~\bibnamefont
  {Kinosita}, \bibfnamefont {Jr.}},\ }\href@noop {} {\bibfield  {journal}
  {\bibinfo  {journal} {Cell}\ }\textbf {\bibinfo {volume} {130}},\ \bibinfo
  {pages} {309} (\bibinfo {year} {2007})}\BibitemShut {NoStop}%
\bibitem [{\citenamefont {Watanabe}\ and\ \citenamefont
  {Noji}(2014)}]{Watanabe2014}%
  \BibitemOpen
  \bibfield  {author} {\bibinfo {author} {\bibfnamefont {R.}~\bibnamefont
  {Watanabe}}\ and\ \bibinfo {author} {\bibfnamefont {H.}~\bibnamefont
  {Noji}},\ }\href@noop {} {\bibfield  {journal} {\bibinfo  {journal} {Nature
  Comm.}\ }\textbf {\bibinfo {volume} {5}} (\bibinfo {year}
  {2014})}\BibitemShut {NoStop}%
\end{thebibliography}
\end{document}